\documentclass[superscriptaddress,aps,prx,twocolumn,showpacs,nofootinbib, longbibliography, notitlepage,floatfix]{revtex4-2}
\usepackage{etex}
\usepackage{amsthm}
\usepackage{amssymb}
\usepackage{url}
\usepackage[colorlinks=true,citecolor=blue,urlcolor=blue]{hyperref}
\usepackage[pdftex]{graphicx}
\usepackage{newtxtext,newtxmath}
\usepackage{braket}
\usepackage{color}
\usepackage{natbib}
\usepackage{amsmath,blkarray}
\usepackage{mathtools}
\usepackage[normalem]{ulem}
\usepackage{latexsym}
\usepackage{tabularx, booktabs}
\usepackage{graphics,epstopdf}
\usepackage{float}
\usepackage{tikz}
\usetikzlibrary{quantikz}
\usepackage{color,soul}
\usepackage{lineno}
\usepackage{comment}    

\usepackage{multirow}
\usepackage{appendix}

\begin{document}
\title{Nonadiabatic Self-Healing of Trotter Errors in Digitized Counterdiabatic Dynamics}%

\author{Mara Vizzuso}
\email{mara.vizzuso@unina.it}
\affiliation{Dipartimento di Fisica ``E. Pancini'', Universit\`a degli Studi di Napoli ``Federico II'', Complesso Universitario M. S. Angelo, via Cintia 21, 80126, Napoli, Italy}

\author{Gianluca Passarelli}
\affiliation{Dipartimento di Fisica ``E. Pancini'', Universit\`a degli Studi di Napoli ``Federico II'', Complesso Universitario M. S. Angelo, via Cintia 21, 80126, Napoli, Italy}

\author{Giovanni Cantele}
\affiliation{Dipartimento di Fisica ``E. Pancini'', Universit\`a degli Studi di Napoli ``Federico II'', Complesso Universitario M. S. Angelo, via Cintia 21, 80126, Napoli, Italy}

\author{Procolo Lucignano}
\affiliation{Dipartimento di Fisica ``E. Pancini'', Universit\`a degli Studi di Napoli ``Federico II'', Complesso Universitario M. S. Angelo, via Cintia 21, 80126, Napoli, Italy}

\author{Xi Chen}
\affiliation{Instituto de Ciencia de Materiales de Madrid (CSIC), Cantoblanco, E-28049 Madrid, Spain}

\author{Koushik Paul}
\email{koushikpal09@gmail.com}
\affiliation{Department of Physical Chemistry, University of the Basque Country UPV/EHU, Apartado 644, 48080 Bilbao, Spain}
\affiliation{EHU Quantum Center, University of the Basque Country UPV/EHU, Barrio Sarriena, s/n, 48940 Leioa, Spain}

\date{\today}

\begin{abstract}

Trotter errors in digitized quantum dynamics arise from approximating time-ordered evolution under noncommuting Hamiltonian terms with a product formula. In the adiabatic regime, such errors are known to exhibit long-time self-healing [Phys. Rev. Lett. \textbf{131}, 060602 (2023)], where discretization effects are effectively suppressed. Here we show that self-healing persists at finite evolution times once nonadiabatic errors induced by finite-speed ramps are compensated. Using counterdiabatic driving to cancel diabatic transitions and isolate discretization effects, we study both noninteracting and interacting spin models and characterize the finite-time scaling with the Trotter steps and the total evolution time. In the instantaneous eigenbasis of the driven (gapped) Hamiltonian, the leading digital error maps to an effective harmonic perturbation whose dominant Fourier component yields an analytic upper bound on the finite-time Trotter error and reveals the phase-cancellation mechanism underlying self-healing. Our results establish finite-time self-healing as a generic feature of digitized counterdiabatic protocols, clarify its mechanism beyond the long-time adiabatic limit, and provide practical guidance for high-fidelity state preparation on gate-based quantum processors.
\end{abstract}
\maketitle

\section{Introduction}\label{sec:introduction}

One of the most promising applications of quantum computing is the simulation of physical and chemical systems~\cite{feynman2018,scherer2017,Kassal2011,Tacchino2020}, which generally requires preparing quantum states evolved to arbitrary times~\cite{Georgescu2014,Buluta2009,Daley2022,Trabesinger2012}. Quantum simulation strategies can be broadly grouped into analog quantum simulation~\cite{hangleiter2022,Daley2023,MacDonell2021} and digital quantum simulation (DQS)~\cite{Lloyd1996,Fauseweh2024,Lanyon2011}, while classical methods provide an important baseline but face intrinsic limitations due to the exponential growth of Hilbert space~\cite{Hairer2006}. Analog simulators emulate many-body dynamics by engineering effective Hamiltonians in controllable platforms such as cold atoms or superconducting circuits. In contrast, DQS encodes the target unitary as a sequence of elementary gates and approximates continuous-time evolution via product formulas such as the Trotter--Suzuki decomposition~\cite{DeRaedt1983,Somma2016}, thereby enabling the simulation of a broad class of spin Hamiltonians on gate-based quantum processors~\cite{Tacchino2019,Dalmonte2016}.

Groundstate preparation is a central task in quantum simulation~\cite{Lin2020}, with broad relevance across condensed matter physics~\cite{Wang2024,Zheng2022}, quantum chemistry~\cite{Motta2023}, quantum information science~\cite{Jaeger2007,VEDRAL19981}, and quantum optimization~\cite{ko2025,Gunther2025}. Many static properties of many-body systems, including phases of matter and correlation functions, are encoded in their groundstates, motivating efficient preparation algorithms~\cite{Poulin2009,Ge2019,Lin2020,Zhang2022,He2022,Solyom1984,Stanisic2022}. Alternative approaches include quantum cooling protocols, where an auxiliary register (e.g., ancillas with measurement-and-reset) is used to extract energy from the target system~\cite{Puente2024,Finkelstein2024}. Furthermore, the development of noisy intermediate-scale quantum (NISQ) hardware~\cite{Preskill2018,Lau2022,Bonde2024} has intensified efforts to explore near-term groundstate preparation methods compatible with finite depth and noise~\cite{Cerezo2021,Bharti2022}.

A widely studied approach to groundstate preparation is adiabatic state preparation (ASP)~\cite{Albash2018}, in which the system evolves under a time-dependent Hamiltonian interpolating from an initial Hamiltonian $H_i$, whose groundstate is easy to prepare, to a target Hamiltonian $H_\mathrm{f}$. ASP has been applied across quantum computation~\cite{Aharonov2008,Albash2018,Hamma2008}, linear algebra~\cite{Subaifmmode2019,Costa2022,An2022}, combinatorial optimization~\cite{farhi2000,Cugini2025}, and quantum simulation~\cite{Jordan2012,Aspuru-Guzik2005,Lee2023,Venuti2017}. Its main limitation is the long runtime required by the adiabatic theorem in the presence of small spectral gaps~\cite{Kato1950,Jansen2007}.
When ASP is implemented on gate-based quantum processors, two error sources typically dominate: diabatic errors due to finite runtime (non-adiabatic transitions)~\cite{Zener1932}, and digital errors due to Trotterization~\cite{Trotter1959,Suzuki1976,Lloyd1996,lu2025}. For time-independent Hamiltonians, product-formula errors accumulate with time: for a first-order decomposition one typically finds a state-preparation infidelity scaling as $\mathcal{O}(t^{2}\Delta t^{2})$~\cite{Childs2021}, where $\Delta t$ is the Trotter step and $t$ the evolution time. These scaling laws highlight the trade-off between accuracy and circuit depth that is central to near-term implementations.
Remarkably, for ASP a self-healing suppression of Trotter errors has been observed~\cite{Kovalsky2023}, where the final state converges to the adiabatic target despite noticeable deviations during the evolution. For smooth schedules and sufficiently small step sizes, e.g.\ $\Delta t=\mathcal{O}\!\left(1/\max_t\|H(t)\|\right)$ with $\|\cdot\|$ the operator (spectral) norm, the final infidelity admits an asymptotic bound of the form $\mathcal{O}(T^{-2}\Delta t^{2})+\mathcal{O}(T^{-2})$, where $T$ is the total runtime and the two terms capture discretization and diabatic contributions, respectively~\cite{Kovalsky2023}. While both terms vanish as $T\to\infty$, at finite $T$ the two error sources can interfere and obscure the intrinsic scaling of the digital error, especially for coarse Trotter steps or schedules inducing rapid spectral variations. Clarifying whether self-healing persists at finite $T$ once diabatic contributions are removed is therefore essential for realistic, depth-limited implementations.

To isolate intrinsic finite-time Trotter errors, a natural strategy is to suppress the diabatic contribution to the evolution. Several approaches can achieve this goal, among which shortcuts to adiabaticity provide a systematic and controllable framework~\cite{Guery2019,TORRONTEGUI2013}. Counterdiabatic (CD) driving~\cite{Demirplak2003,Demirplak2005,Berry2009,DelCampo2013} is a central shortcut-to-adiabaticity technique, widely used in quantum control~\cite{Petiziol2024,Iram2020}, state transfer~\cite{Li2016,Feng2017}, and optimization algorithms~\cite{Vizzuso2024,Vizzuso20241,Hegade2021,Chandarana2022}. It augments the reference Hamiltonian with a (typically) time-dependent auxiliary term designed to cancel non-adiabatic transitions and enforce tracking of instantaneous eigenstates. Exact CD terms require full spectral information and are generally impractical for many-body systems; nevertheless, approximate and variational schemes, including local ans\"atze~\cite{Sels2017} and nested-commutator constructions~\cite{Claeys2019}, can substantially suppress diabatic errors at moderate runtimes. In the present work, CD driving serves as a diagnostic/control knob to separate diabatic and digital errors at finite $T$.

In this work, we study Trotterized adiabatic evolution in the presence of CD potentials with the goal of isolating discretization-induced errors from diabatic transitions at finite runtime. We show that, once diabatic contributions are suppressed, the residual digital error acquires a bounded oscillatory structure in time rather than accumulating monotonically. We further derive an effective harmonic description whose dominant Fourier component yields an analytic upper bound on the finite-time Trotter error and exposes the phase-cancellation mechanism underlying self-healing. We validate these predictions in both noninteracting and interacting spin models including Ising chains and $p$-spin model, demonstrating the robustness of the mechanism across distinct dynamical regimes.

The remainder of the paper is organized as follows. In Sec.~\ref{sec:Preliminaries}, we introduce the general framework for analyzing Trotter errors in both adiabatic and CD-assisted evolutions, including known error bounds and a review of exact and approximate CD constructions. In Sec.~\ref{sec:analytical-description}, we present our main theoretical results, deriving analytical expressions for finite-time digital errors and their oscillatory structure. The spin models are described in Sec.~\ref{sec:model}, followed by numerical simulations and benchmarks in Sec.~\ref{sec:Examples}. Finally, Sec.~\ref{sec:conclusions} summarizes our findings and discusses implications for digitized quantum simulation and state preparation on gate-based quantum processors.

\section{Preliminaries}\label{sec:Preliminaries}

We begin by establishing the key theoretical elements underlying our study. This section summarizes (i) standard bounds on diabatic (non-adiabatic) errors at finite evolution times and (ii) digitization errors arising from product-formula approximations, and introduces CD driving as a strategy to suppress non-adiabatic transitions. These considerations lay the groundwork for analyzing CD-assisted digitized adiabatic evolution and the emergence of self-healing behavior. Throughout this paper, we set $\hbar=1$. In this paper, we denote the real evolved state by $\ket{\psi(t)}$, the state evolved under digitized evolution by $\ket{\psi_\text{T}(t)}$, and the instantaneous eigenstate by $\ket{\phi(t)}$. These states can be defined for the entire interval $t\in[0,T]$.

\subsection{Errors and bounds}\label{sec:Errors-and-bounds}

In general, the dynamics generated by a time-dependent Hamiltonian is analytically intractable. Consider a system initialized at $t=0$ and evolved up to a final time $t=T$ according to the time-dependent Schr\"odinger equation
\begin{equation}
   i \partial_t \ket{\psi(t)} = H(t)\ket{\psi(t)}.
   \label{eq:shrodinger}
\end{equation}
Formally, the corresponding time-evolution operator is
\begin{equation}
    U(t)=\mathcal{T}\exp\!\left[-i \int_0^t dt' \,H(t')\right],
    \label{eq:real-evolution-operator}
\end{equation}
where $\mathcal{T}$ denotes time ordering operator. Understanding such time evolution is essential in quantum computation and control, as it determines how quantum states transform under designed protocols. However, in most cases, solving the Schr\"odinger equation exactly is infeasible.

A common simplification is provided by the adiabatic theorem~\cite{Albash2018}: if $H(t)$ is sufficiently smooth, the energy spectrum remains gapped, and the Hamiltonian is varied sufficiently slowly compared with the inverse gap scale, then a system initialized in an instantaneous eigenstate $\ket{\phi_n(0)}$ of $H(0)$ approximately follows the corresponding instantaneous eigenstate $\ket{\phi_n(t)}$ of $H(t)$ up to a phase factor. For long-time evolutions, $T\to\infty$, the deviation from the adiabatic trajectory vanishes, whereas for large but finite $T$ the diabatic error scales as
\begin{equation}
    \bigl\|\ket{\psi(t)}-\ket{\psi_{\rm ad}(t)}\bigr\|\sim \mathcal{O}\!\left(\frac{1}{T}\right),
    \label{eq:adiabatic-error}
\end{equation}
with a prefactor that depends on $\max_t\|\dot H(t)\|$ and the minimum spectral gap $\Delta_{\min}$, where $\ket{\psi_{\rm ad}(t)}$ denotes the ideal adiabatic state in the slow-driving limit~\cite{Aharonov2007}.

To simulate such dynamics on a gate-based quantum processor, one discretizes time, leading to a digitized (Trotterized) approximation. For concreteness, we consider an interpolation of the form $H(t)=A(t)H_{\rm i}+B(t)H_{\rm f}$ and divide the interval $[0,T]$ into $M$ uniform steps of size $\Delta t=T/M$. We will primarily analyze stroboscopic times $t_m=m\Delta t$ with integer $m\in\{0,1,\dots,M\}$; for a general time $t$ one may take $m=\lfloor t/\Delta t\rfloor$.
The exact propagator at $t_m$ is then approximated by the first-order product formula
\begin{equation}
    U_{\rm T}(t) =
    \prod_{k=0}^{m-1} e^{-i \Delta t\,A(t_k)\,H_{\rm i}}\,
    e^{-i \Delta t\,B(t_k)\,H_{\rm f}},
    \qquad t_k=k\Delta t+ \Delta t/2,
    \label{eq:digitalized-evolution}
\end{equation}
with $U_{\rm T}(0)=\openone$. This first-order Trotter--Suzuki decomposition incurs a local error per step of order
$\mathcal{O}\!\left((\Delta t)^2 \,\|[H_{\rm i},H_{\rm f}]\|\right)$
(up to prefactors depending on $A,B$), where $\|\cdot\|$ denotes the operator (spectral) norm. In the limit $\Delta t\to 0$, $U_{\rm T}(t_m)$ approaches the continuous propagator $U(t_m)$.

To quantify digitization errors, we use the (state-dependent) digital infidelity
\begin{equation}
    \mathcal{I}
    (t)=1-\bigl|\langle\psi(t)|\psi_{\rm T}(t)\rangle\bigr|^2,
    \label{eq:digital-infidelity}
\end{equation}
where $|\psi(t)\rangle = U(t_m)|\psi(0)\rangle$ and $|\psi_{\rm T}(t_m)\rangle = U_{\rm T}(t_m)|\psi(0)\rangle$.
When our goal is groundstate preparation we also consider the final groundstate infidelity
\begin{equation}
    \mathcal{I}_{\rm gs}(t)=1-\bigl|\langle\phi_0(t)|\psi_{\rm T}(t)\rangle\bigr|^2,
    \label{eq:gs-infidelity}
\end{equation}
where $|\phi_0(t)\rangle$ is the groundstate of the target Hamiltonian $H(t)$ and $|\psi_{\rm T}(t)\rangle$ is the digitized state.  The quantity $\mathcal{I}(t)$ isolates discretization errors by comparing the digitized evolution to the exact continuous evolution under the same driving, whereas $\mathcal{I}_{\rm gs}(t)$ measures the overall state-preparation error with respect to the target groundstate and therefore includes both diabatic and digital contributions (unless diabatic transitions are suppressed, e.g., by CD driving). If the Hamiltonian $H(t)$ includes the exact counterdiabatic (CD) potential and the initial state is the groundstate of the initial Hamiltonian, then, by definition, the system remains in the instantaneous groundstate throughout the evolution, i.e., $\ket{\psi(t)} \equiv \ket{\phi_0(t)}$ for all $t$, consequently $\mathcal{I}_\text{gs}(t) \equiv \mathcal{I}(t)$. Thus, throughout this work, analyzing $\mathcal{I}(t)$ in the exact CD regime is equivalent to studying the groundstate infidelity $\mathcal{I}_\text{gs}(t)$.


Naively, one might expect digitization errors to accumulate monotonically with the number of steps. However, as shown in Ref.~\cite{Kovalsky2023}, for sufficiently large total runtime $T$ interference among successive Trotter steps can partially cancel these errors, a phenomenon known as self-healing. In particular, in the long-time regime one finds a scaling of the digital contribution as $\mathcal{O}(T^{-2}\Delta t^{2})$ for first-order Trotterization~\cite{Kovalsky2023}. In the following, we go beyond this asymptotic regime and demonstrate that self-healing can persist even at finite evolution times once diabatic contributions are properly suppressed.

\subsection{Counterdiabatic driving and variational gauge potentials} \label{sec:Counterdiabatic-potential}

To investigate this behavior, it is necessary to suppress the non-adiabatic (diabatic) errors that arise at finite evolution times. To this end, we incorporate an auxiliary CD term into the time-dependent Hamiltonian $H(t)$, designed to cancel transitions between instantaneous eigenstates and enforce adiabatic tracking.

We parametrize the driving by a schedule $\lambda(t)\in[0,1]$ and write the reference Hamiltonian as
$H(t)=H(\lambda(t))=A(\lambda(t))H_{\rm i}+B(\lambda(t))H_{\rm f}$. For the results presented below we take smooth schedules; when needed we choose boundary conditions such that $\dot\lambda(0)=\dot\lambda(T)=0$ so that the CD term vanishes at the endpoints. The total Hamiltonian under CD driving reads
\begin{equation}
H_{\rm tot}(t)=H(t)+H_{\rm CD}(t),
\label{eq:real-CD-evolution}
\end{equation}
where the CD term is written in terms of the adiabatic gauge potential (AGP) as
$H_{\rm CD}(t)=\dot{\lambda}(t)\,\mathcal{A}_{\lambda}(\lambda(t))$.
For nondegenerate spectra, the AGP satisfies~\cite{Berry2009}
\begin{equation}
\bra{\phi_m(\lambda)}\mathcal{A}_\lambda\ket{\phi_n(\lambda)}
= -i \frac{\bra{\phi_m(\lambda)}\partial_\lambda H(\lambda)\ket{\phi_n(\lambda)}}{E_m(\lambda)-E_n(\lambda)},
\label{eq:CD-real-potential}
\end{equation}
with $\{\ket{\phi_n(\lambda)}\}$ the instantaneous eigenstates of $H(\lambda)$ and $\{E_n(\lambda)\}$ the corresponding eigenenergies. The diagonal elements of $\mathcal{A}_\lambda$ can be fixed by a gauge choice and do not affect transition suppression.

For small systems, $\mathcal{A}_\lambda$ can be constructed explicitly from the instantaneous spectrum, yielding an (essentially) exact CD correction that eliminates diabatic transitions. For larger systems, evaluating $\mathcal{A}_\lambda$ exactly is computationally intractable, and we therefore adopt a variational approximation~\cite{Sels2017,Claeys2019,Passarelli2020}. Specifically, we approximate the AGP by a truncated nested-commutator expansion,
\begin{equation}
\mathcal{A}=i  \sum_{k=1}^l\alpha_k\,O_{2k-1},
\label{eq:variational-ansatz-CD-potential}
\end{equation}
where $O_0=\partial_\lambda H$, $ 
O_k=[H,O_{k-1}]\ (k\ge 1)$, and the coefficients $\vec{\alpha}=\{\alpha_k\}$ are determined by minimizing the action~\cite{Sels2017}
\begin{equation}
\mathcal{S}(\mathcal{A}_\lambda)=\mathrm{Tr}\!\left[G_\lambda^\dagger(\mathcal{A}_\lambda)\,G_\lambda(\mathcal{A}_\lambda)\right],
\label{eq:action-formal}
\end{equation}
where $G_\lambda(\mathcal{A}_\lambda)\equiv \partial_\lambda H+i [\mathcal{A}_\lambda,H]$.
For the ans\"atz~\eqref{eq:variational-ansatz-CD-potential}, one has
$G_\lambda(\mathcal{A})=O_0-\sum_{k=1}^l \alpha_k\,O_{2k}$,
and therefore the action reduces to a quadratic form
\begin{equation}
S_l(\vec{\alpha}) \equiv \mathcal{S}(\mathcal{A}_\lambda)
= A + 2\,\vec{B}\cdot\vec{\alpha} + \vec{\alpha}^{\,T}\underline{C}\,\vec{\alpha},
\label{eq:system-action}
\end{equation}
with
\begin{equation}
A=\mathrm{Tr}[O_0^2],\
B_i=-\mathrm{Tr}[O_0\,O_{2i}],\
C_{ij}=\mathrm{Tr}[O_{2i}\,O_{2j}].
\label{eq:ABC-def}
\end{equation}
Minimizing~\eqref{eq:system-action} yields the linear system $\underline{C}\,\vec{\alpha}=-\vec{B}$, whose solution defines the variational AGP $\mathcal{A}_\lambda^{(l)}$ and hence an approximate CD term $H_{\rm CD}(t)=\dot\lambda(t)\mathcal{A}_\lambda^{(l)}$.
For notational simplicity, we fix a truncation order $l$ in what follows.

Both exact and variational CD driving can be incorporated into the digitized framework by augmenting each Trotter step with the CD evolution.
Using the same stroboscopic times $t_m=m\Delta t$ as in Sec.~\ref{sec:Errors-and-bounds}, we write
\begin{equation}\label{eq:CD-digitalized-evolution}
U_{\rm T}^{\rm CD}(t_m)
\simeq \prod_{k=0}^{m-1}
e^{-i \Delta t\,A(t_k)\,H_{\rm i}}
e^{-i \Delta t\,B(t_k)\,H_{\rm f}}
e^{-i \Delta t\,H_{\rm CD}(t_k)} .
\end{equation}
On hardware, the short-time unitaries in Eq.~\eqref{eq:CD-digitalized-evolution} can be compiled into elementary gates by decomposing the generators into available native interactions (e.g., Pauli strings)~\cite{Chen2022}. Furthermore, Ref.~\cite{mizuta2025} provides a systematic framework to control the scaling of Trotter errors with the system size $N$, by deriving explicit bounds in terms of nested commutators of the Hamiltonian. In the analysis below, we treat the factors in Eq.~\eqref{eq:CD-digitalized-evolution} as the elementary digitization blocks to isolate the interplay between diabatic suppression and product-formula errors; additional compilation/Trotterization of $H_{\rm CD}$ itself can be included as an extra (implementation-dependent) error source.


\section{Analytical Description}\label{sec:analytical-description}

Although various metrics exist to quantify product-formula Trotter errors~\cite{DeRaedt1983},  we adopt the (state-dependent) infidelity between the exact continuous-time state and the corresponding digitized state as our figure of merit, see Eq. (\ref{eq:digital-infidelity}). This choice is motivated by the fact that, when evaluated against the target adiabatic trajectory, the infidelity captures the combined influence of diabatic (finite-$T$) effects and discretization (Trotter) effects. When counterdiabatic (CD) driving is included, we analogously adapt the quantity in Eq.~\eqref{eq:digital-infidelity} to the evolution generated by a (gapped) Hamiltonian containing an exact or variational CD potential. In this case, the state $\ket{\psi(t)}$ evolves under the Hamiltonian in Eq.~\eqref{eq:real-CD-evolution}, while the digitalized state is generated by the operators in Eq.~\eqref{eq:CD-digitalized-evolution}. The CD term $H_\text{CD}(t)$ may be either exact or approximated.

We emphasize that our simulations are performed in \textit{Python} using the \textit{QuTiP} library. In particular, the time-dependent Schr\"odinger equation (see Eq.~\eqref{eq:shrodinger}) is solved numerically using the \textit{sesolve} function~\cite{qutip_sesolve}. This solver integrates the equation by means of standard Ordinary Differential Equation (ODE) methods, specifically adaptive-step integrators based on the ZVODE algorithm. The numerical accuracy is controlled through relative and absolute tolerances, set to values on the order of $10^{-10}$. Within these settings, the method provides an accurate approximation of the quantum state $\ket{\psi(t)}$ over the entire time evolution.

By analogy with time-independent Hamiltonians, one expects $\mathcal{I}(t) \sim \mathcal{O}(t^{2}\Delta t^{2})$ for first-order Trotterization, and the error generally grows with the evolution time $t$, since longer evolutions require a larger sequence of elementary unitaries (see~Eqs.~\eqref{eq:digitalized-evolution} -~\eqref{eq:CD-digitalized-evolution}). The precise dependence of digitization errors on the total runtime $T$ and discretization step $\Delta t$ for time-dependent Hamiltonians, however, is more subtle. Previous work~\cite{Kovalsky2023} demonstrated a self-healing behavior in the long-time regime $T\to\infty$, where interference among successive Trotter steps suppresses the net discretization error and the leading contribution scales as $\mathcal{O}(T^{-2}\Delta t^{2})$.

This raises the question of whether self-healing behavior is universal, i.e., whether the residual digitization error is predominantly determined by the discretization step $\Delta t$, while diabatic errors of order $\mathcal{O}(1/T)$ dominate only when the evolution is insufficiently adiabatic. If so, correcting non-adiabaticity~\cite{Comparat2009} could preserve adiabatic features of the evolution and enable self-healing even at finite $T$.
It can be shown that the final groundstate infidelity at $t=T$ can be bounded in terms of instantaneous populations in the eigenbasis of the reference Hamiltonian~\cite{Kovalsky2023}.
In particular, at $\lambda(T)=1$ one has
\begin{equation}
\mathcal{I}(T)
\le \sum_{i \neq 0} \left| P_i(T,1) \right|^2,
\label{eq:dis-infidelity}
\end{equation}
where the sum excludes $i=0$ because the system is initialized in the groundstate and the infidelity is therefore associated with population transferred to excited levels. To define the overlaps appearing in Eq.~\eqref{eq:dis-infidelity}, we work in the instantaneous eigenbasis of the reference Hamiltonian $H(\lambda)$ 
and introduce
\begin{equation}
P_i(t,\lambda)\equiv \langle \phi_i(\lambda)|\psi(t)\rangle .
\label{eq:Pi-definition}
\end{equation}
Here $|\psi(t)\rangle$ denotes the evolving state; in the remainder of this section, unless stated otherwise, it should be understood as the digitized state generated by the chosen product formula (with or without CD), so that $P_i(t,\lambda)$ tracks the stroboscopic dynamics of the digitized overlaps in the instantaneous eigenbasis of $H(\lambda)$.

To quantify the error arising from the Trotterized time evolution, it is therefore necessary to analyze the discrete-time propagation of these overlaps.
For stroboscopic times $t_m=m\Delta t$, the evolution of $P_i$ over a single Trotter step can be expressed as (cf.~Appendix~\ref{app:SRQ})
\begin{equation}
P_i(t_{m+1},\lambda)
=\sum_{j,k}G_{kj}(\lambda,\Delta t)\,P_j(t_m,\lambda)\,M_{ik}(\lambda).
\label{eq:Pi-relation}
\end{equation}
Here, $G_{kj}(\lambda,\Delta t) \equiv 
\langle \phi_k(\lambda) | U^{\rm CD}_{\rm T}(\lambda,\Delta t) | \phi_j(\lambda) \rangle$, while  $M_{ik}(\lambda) \equiv \langle \phi_i(\lambda+\Delta\lambda) | \phi_k(\lambda) \rangle$. The explicit dependence of both $G_{kj}(\lambda,\Delta t)$ and $M_{ik}(\lambda)$ on $\Delta t$ and $T$, arising from the structure of the digitized evolution operator (e.g.\ $U_{\rm T}^{\rm CD}$) and from the adiabatic-basis transformation along the schedule, is derived in Appendix~\ref{app:SRQ}.
It is worth noting that, by expressing the evolution of $P_i$ in this form, distinct contributions to the total error can be separated: $G_{kj}$ captures discretization effects from the product formula, while $M_{ik}$ captures the basis change associated with the instantaneous eigenbasis.
Eq.~\eqref{eq:Pi-relation} can be further decomposed in terms of three auxiliary functions, $\mathcal{S}_{ij}$, $\mathcal{R}_{ij}$, and $\mathcal{Q}_{ij}$ (see Appendix~\ref{app:SRQ}), which encode, respectively, the contributions associated with basis-change/adiabatic couplings and with digitization-induced mixing.
Once the dynamics are expressed in terms of these quantities, discrete differentiation of Eq.~\eqref{eq:Pi-relation} yields a Schr\"odinger-like equation for the overlap amplitudes,
\begin{align}
i \frac{\partial P_i(t, \lambda)}{\partial t}
&\simeq E_i(\lambda)\,P_i(t, \lambda)\notag\\
&\quad + i \sum_{j\neq i}\left[\mathcal{Q}_{ij}+\mathcal{S}_{ij}+\mathcal{R}_{ij}\right] P_j(t, \lambda),
\label{eq:schrodinger-Pi-complete}
\end{align}
where $E_i(\lambda)$ is the instantaneous eigenenergy of the reference Hamiltonian $H(\lambda)$ associated with $|\phi_i(\lambda)\rangle$.
The first (diagonal) term describes the dynamical phase accumulation in the instantaneous eigenbasis, while the off-diagonal terms describe mixing between different instantaneous eigenstates induced by finite-$T$ and/or digitization effects.
For conventional adiabatic evolution (without CD), the terms $\mathcal{S}_{ij}$ and $\mathcal{Q}_{ij}$ encode non-adiabatic couplings arising from the basis change along the schedule and scale as $\mathcal{O}(1/T)$, whereas $\mathcal{R}_{ij}$ originates from the product-formula discretization and scales with the Trotter step $\Delta t$~\cite{Kovalsky2023}. As discussed in Appendix~\ref{app:SRQ}, including the CD potential suppresses the non-adiabatic contributions, yielding $\mathcal{S}_{ij}\to 0$ and $\mathcal{Q}_{ij}\to 0$, while $\mathcal{R}_{ij}$ remains and scales as $\mathcal{O}(\Delta t)$. Consequently, the dominant residual off-diagonal contribution in Eq.~\eqref{eq:schrodinger-Pi-complete} is governed by $\mathcal{R}_{ij}$, highlighting that, in the instantaneous eigenbasis of $H(\lambda)$, the remaining error in CD-assisted digitized evolution is primarily set by finite discretization.

If the schedule $\lambda(t)$ (e.g., $\lambda(t)=\sin^2(\pi t/2T)$) satisfies the endpoint-smooth condition
$\dot{\lambda}(0)=\dot{\lambda}(T)=0$,
then the CD term, $H_{\rm CD}(t)=\dot{\lambda}(t)\mathcal{A}_\lambda$, vanishes at the protocol boundaries,
$H_{\rm CD}(0)=H_{\rm CD}(T)=0$, such that
$H_{\rm tot}(0)=H(0)$ and $H_{\rm tot}(T)=H(T)$.
In this case, the digitized evolution at $\lambda=0$ and $\lambda=1$ reduces to a purely diagonal phase accumulation in the instantaneous eigenbasis of the reference Hamiltonian $H(\lambda)$, and the discretization-induced off-diagonal mixing term obeys
\begin{equation}
\mathcal{R}_{ij}(0,\Delta t)=\mathcal{R}_{ij}(1,\Delta t)=0.
\label{eq:Rij-endpoints}
\end{equation}
Therefore, it is natural to expand $\mathcal{R}_{ij}$ in a sine series on $\lambda\in[0,1]$,
\begin{equation}
\mathcal{R}_{ij}(\lambda,\Delta t)=\sum_{q\ge 1} c_q\,R(\Delta t)\,\sin(\pi q\lambda).
\label{eq:harmonic-potential}
\end{equation}
The quantity $R(\Delta t)$ should more properly be written as $R_{ij}(\Delta t)$, since the amplitudes of a single mode depend on the energy levels $i$ and $j$. Indeed, for fixed $i$ and $j$, one can identify an effective harmonic potential describing the overlap between these energy levels. For notational simplicity, we denote $R_{ij}(\Delta t)$ by $R(\Delta t)$. Moreover, independently of the specific values of $i$ and $j$, the scaling behavior $R(\Delta t) \sim \mathcal{O}(\Delta t)$ still holds. In practice, the series is often dominated by a single Fourier component; we denote its index by $\bar q$ and retain only this leading harmonic in the perturbative analysis below.

Starting from Eq.~\eqref{eq:schrodinger-Pi-complete}, the diagonal terms correspond to phase accumulation in the instantaneous eigenbasis of $H(\lambda)$, while the off-diagonal terms induce transitions.
Treating the dominant harmonic in Eq.~\eqref{eq:harmonic-potential} as a weak perturbation, the first-order transition amplitude from the instantaneous groundstate to an excited state $i\neq 0$ can be written as
\begin{equation}
P_i(t,\lambda)
\simeq -\,i \int_0^\lambda d\lambda^{\prime}\,
\frac{R(\Delta t)\,\sin\!\left(\pi \bar{q}\,\lambda^{\prime}\right)}
{\dot{\lambda}\!\left(\lambda^{\prime}\right)}
\, e^{-\,i\varphi_{i0}(\lambda')},
\label{eq:adiabatic-perturbation}
\end{equation}
where the symbol $\simeq$ reflects the fact that, in general, the off-diagonal terms are not described by a single mode, but rather satisfy Eq.~\eqref{eq:harmonic-potential}. The phase is
\begin{equation}
\varphi_{i0}(\lambda')
= \int_0^{\lambda'} d\xi\;
\frac{\Delta_{i0}(\xi)}{\dot{\lambda}(\xi)},
\label{eq:adiabatic-perturbation-phase}
\end{equation}
with $\Delta_{i0}(\lambda)= E_i(\lambda)-E_0(\lambda)$.
Up to this point, we have shown that, once the non-adiabatic couplings are suppressed by CD driving, the residual mixing can be captured as a harmonic perturbation whose overall strength scales as $R(\Delta t)\sim\mathcal{O}(\Delta t)$.
In general, evaluating Eqs.~\eqref{eq:adiabatic-perturbation}-\eqref{eq:adiabatic-perturbation-phase} analytically is challenging for arbitrary schedules. We therefore specialize to the linear schedule $\lambda(t)=t/T$, which yields a closed-form expression and serves as an analytical reference. We emphasize, however, that a strict linear ramp does not satisfy the endpoint-smooth condition, so Eq.~\eqref{eq:Rij-endpoints} is not automatically enforced. When the endpoint-vanishing property is required for the harmonic analysis, it can be restored by adopting a smooth schedule, as discussed in the Supplementary Material.

Specifically, for $\lambda=t/T$ one has $\dot{\lambda}=1/T$, and Eq.~\eqref{eq:adiabatic-perturbation} becomes
\begin{equation}
P_i(t)
\simeq -i  \int_0^t dt^{\prime}\;
e^{-\,i\int_0^{t^\prime}\Delta_{i0}(\tau/T)\,d\tau}
R(\Delta t)\,\sin\!\left(\frac{\pi \bar{q} t^{\prime}}{T}\right)\,
.
\label{eq:adiabatic-perturbation-ramp}
\end{equation}
To expose the interference structure, we approximate the gap as slowly varying over the dominant contribution and take $\Delta_{i0}(1)=\Delta_{i0}$, which yields
\begin{equation}
P_i(t)
\simeq \frac{-\,iR(\Delta t)}{2}\left[
\frac{e^{-\,i(\Delta_{i0}+\pi\bar{q}/T)t}-1}{\Delta_{i0}+\pi\bar{q}/T}
-\frac{e^{-\,i(\Delta_{i0}-\pi\bar{q}/T)t}-1}{\Delta_{i0}-\pi\bar{q}/T}
\right].
\label{eq:P-i-ramp-time}
\end{equation}
Evaluating at $t=T$ gives
\begin{equation}
P_i(T)
\simeq
\frac{-\,iR(\Delta t)}{2}\Bigl(1-(-1)^{\bar{q}}e^{-\,i\Delta_{i0}T}\Bigr)\,
\frac{\pi\bar{q}/T}{\Delta_{i0}^{\,2}-(\pi\bar{q}/T)^2}, 
\label{eq:perturbation-theory-T}
\end{equation}
where $\Delta_{i0}= E_i(1)-E_0(1)$ under the linear-ramp approximation.
The factor $1-(-1)^{\bar{q}}e^{-\,i\Delta_{i0}T}$ encodes the phase-cancellation mechanism underlying self-healing, while the denominator indicates potential resonant enhancement when $\Delta_{i0}\approx \pi\bar{q}/T$.

Away from resonance, i.e.\ when $|\Delta_{i0}\pm\pi\bar{q}/T|$ remains bounded away from zero, and for sufficiently small $R(\Delta t)$, Eq.~\eqref{eq:perturbation-theory-T} implies
\begin{equation}
|P_i(T)|\sim \mathcal{O}\!\left(\frac{R(\Delta t)}{T}\right)
=\mathcal{O}\!\left(\frac{\Delta t}{T}\right),
\end{equation}
up to gap-dependent prefactors. Consequently,
\begin{equation}
\mathcal{I}(T)=\sum_{i\neq 0}|P_i(T)|^2
\sim \mathcal{O}\!\left(T^{-2}\Delta t^{2}\right),
\end{equation}
within the perturbative regime. For very short runtimes or near-resonant conditions, higher-order terms in $R(\Delta t)$ and/or stronger time-dependence of $\Delta_{i0}(\lambda)$ become important, and the first-order description must be supplemented by the higher-order analysis presented in Appendix~\ref{app:SRQ}.

\section{Models}\label{sec:model}

In this section we specify the spin models and protocols used in our numerical benchmarks. We consider three systems composed of $N$ spin-$1/2$ particles and implement an annealing-type interpolation
\begin{equation}
H(t)=A(\lambda(t))\,H_{\rm i}+B(\lambda(t))\,H_{\rm f},\qquad \lambda(t)\in[0,1],
\label{eq:annealing-protocol}
\end{equation}
with $A(\lambda)=1-\lambda$ and $B(\lambda)=\lambda$. Throughout this section we use the linear schedule $\lambda(t)=t/T$ to match our analytical reference calculations.  Here, to validate our theoretical findings, we choose a transverse-field initial Hamiltonian,
\begin{equation}
H_{\rm i}=-2K\,S_x,
\label{eq:hx}
\end{equation}
with $K=1$. The collective spin operators are defined as
\begin{equation}
S_\alpha=\frac{1}{2}\sum_{n=1}^N\sigma_n^{\alpha},\qquad \alpha\in\{x,y,z\},
\label{eq:spin-total}
\end{equation}
where $\sigma_n^{\alpha}$ are Pauli operators acting on qubit $n$.
For the final Hamiltonian $H_{\rm f}$, we consider two classes of interacting target Hamiltonians.
One is fully connected Ising model (collective interaction), we take
\begin{equation}
H_{\rm f}=H_{p}=-\frac{J_p}{N^{p-1}}(2S_z)^p,
\label{eq:hp}
\end{equation}
which for $p=2$ corresponds to a fully connected (infinite-range) Ising model. Unless stated otherwise, we set $J_p=1$ and focus on $p=2$. The other is one-dimensional Ising chain (short-range interaction). 
We study a nearest-neighbor Ising chain with a longitudinal field,
\begin{equation}
H_{\rm f}=H_{ZZ}=-J_Z\sum_{n=1}^{N}\sigma_{n}^{z}\sigma_{n+1}^{z}
+h\sum_{n=1}^{N}\sigma_{n}^{z},
\label{eq:hz}
\end{equation}
where $J_Z$ is the coupling strength and $h$ is a longitudinal field. We assume periodic boundary conditions, $\sigma_{N+1}^{z}\equiv\sigma_{1}^{z}$, unless stated otherwise.

To gain intuition in the absence of interactions, we additionally use a single-qubit benchmark (noninteracting case).
For interacting systems, we simulate the Ising chain~\eqref{eq:hz} at $N=6$ for coupling strengths $J_Z\in\{0.1,0.5,1\}$ (with $h=1$ fixed throughout the simulations), and compare with the fully connected model~\eqref{eq:hp} at $p=2$ for sizes $N\in\{10,30\}$. Also, we employ two types of CD potentials: (i) an exact CD term when the required instantaneous spectral information can be obtained, and (ii) a variational CD approximation when an exact construction is impractical (Sec.~\ref{sec:Counterdiabatic-potential}). Specifically, for the Ising chain we use the exact CD potential, while for the fully connected model we use the variational CD construction. In the latter case we include nested-commutator terms up to truncation order $l=7$.

\begin{figure}
    \hspace{\textwidth}(a)\hspace{0.23\textwidth}(b)\\
    \centering
    \includegraphics[width=0.49\columnwidth]{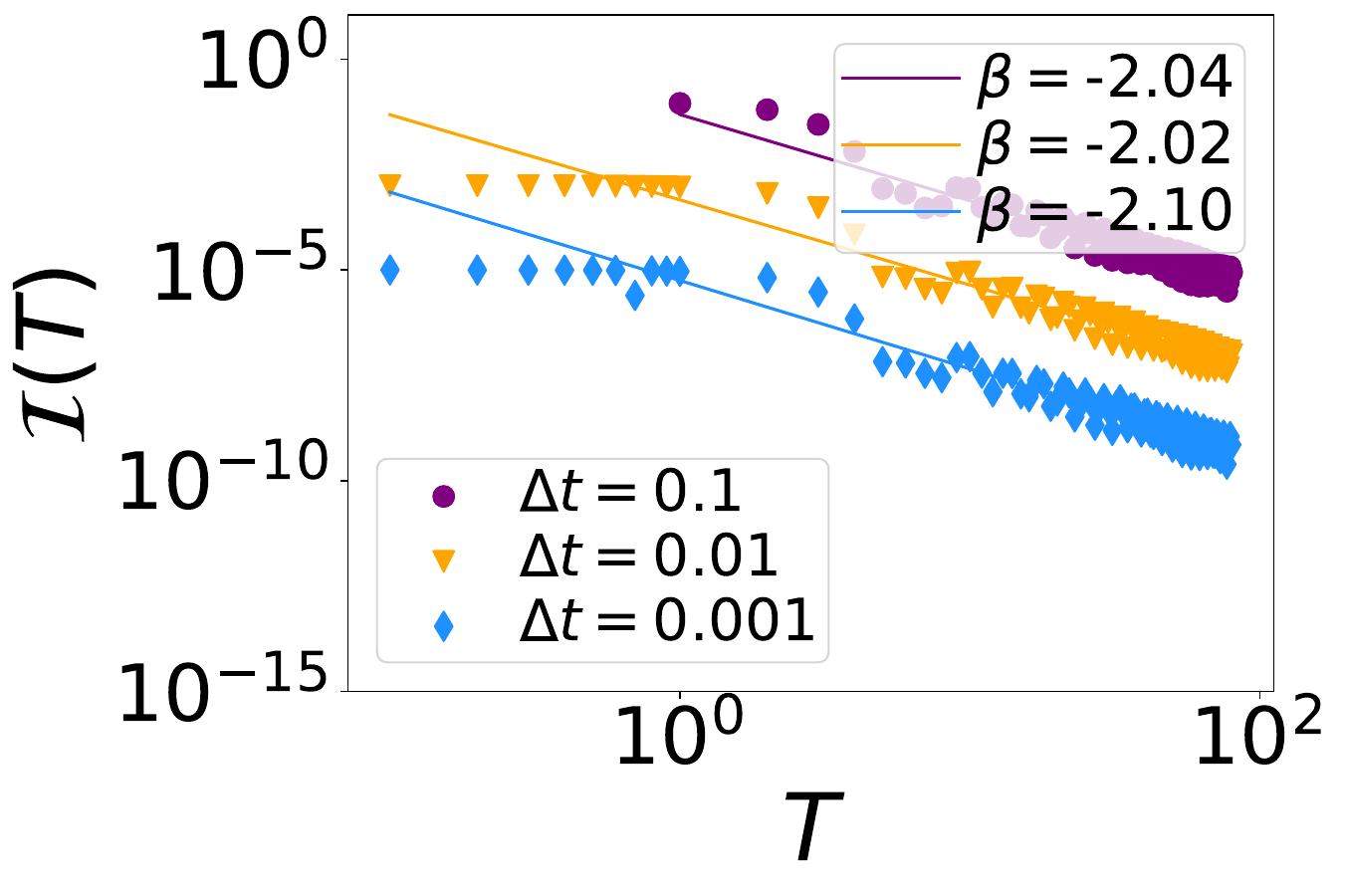}
    \includegraphics[width=0.49\columnwidth]{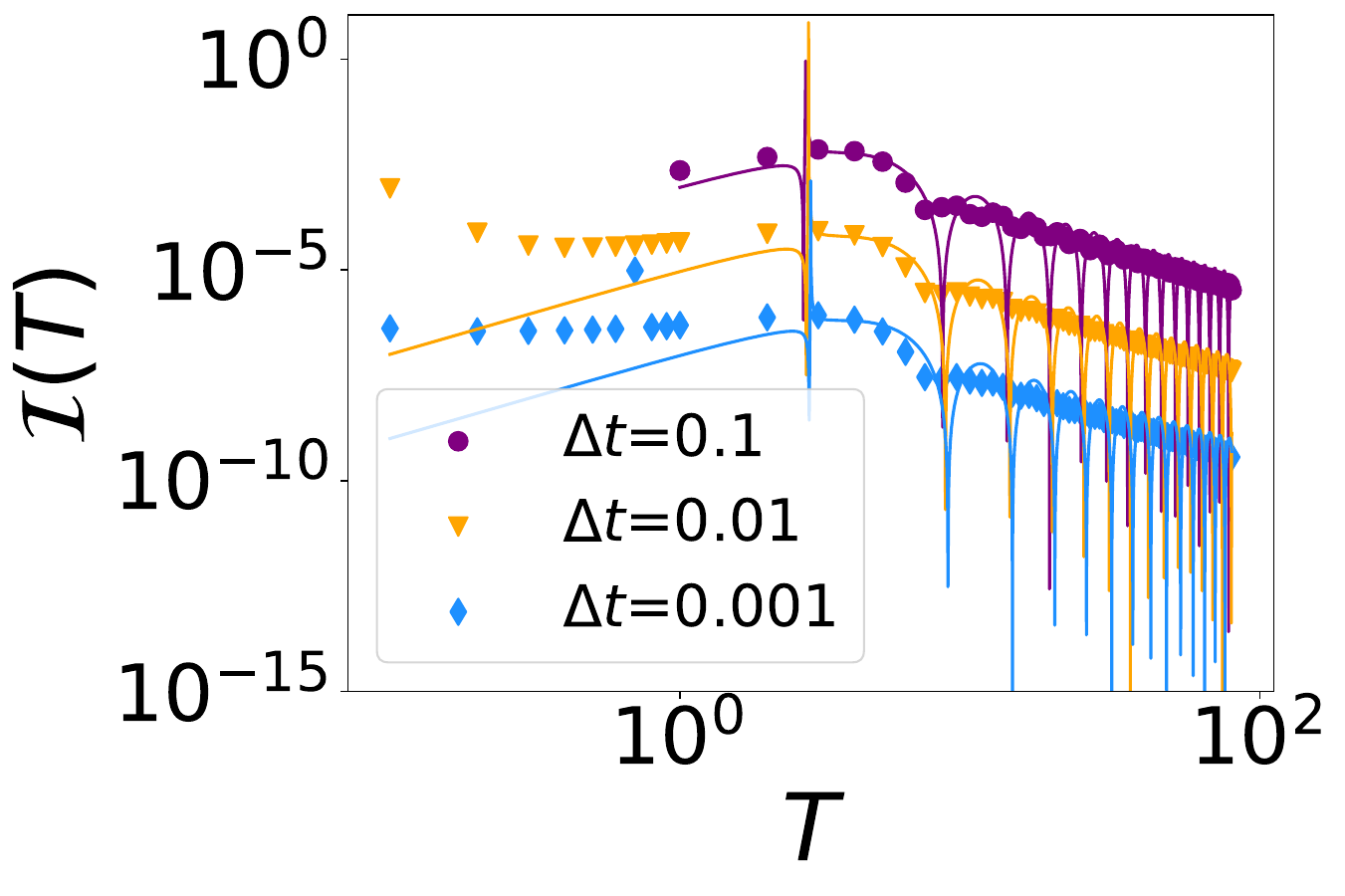}
    \hspace{\textwidth}(c)\hspace{0.23\textwidth}(d)\\
    \centering
    \includegraphics[width=0.5\columnwidth]{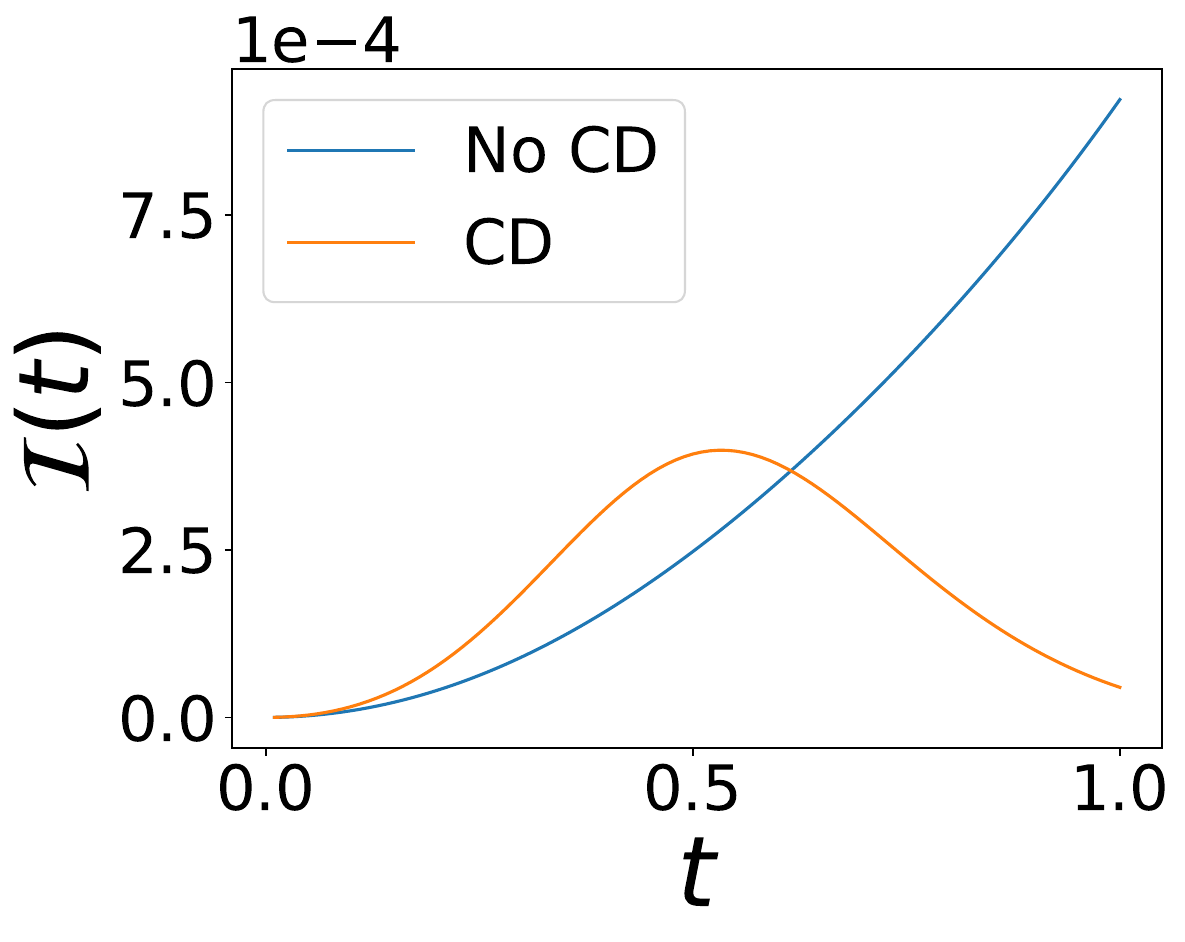}
    \includegraphics[width=0.48\columnwidth]{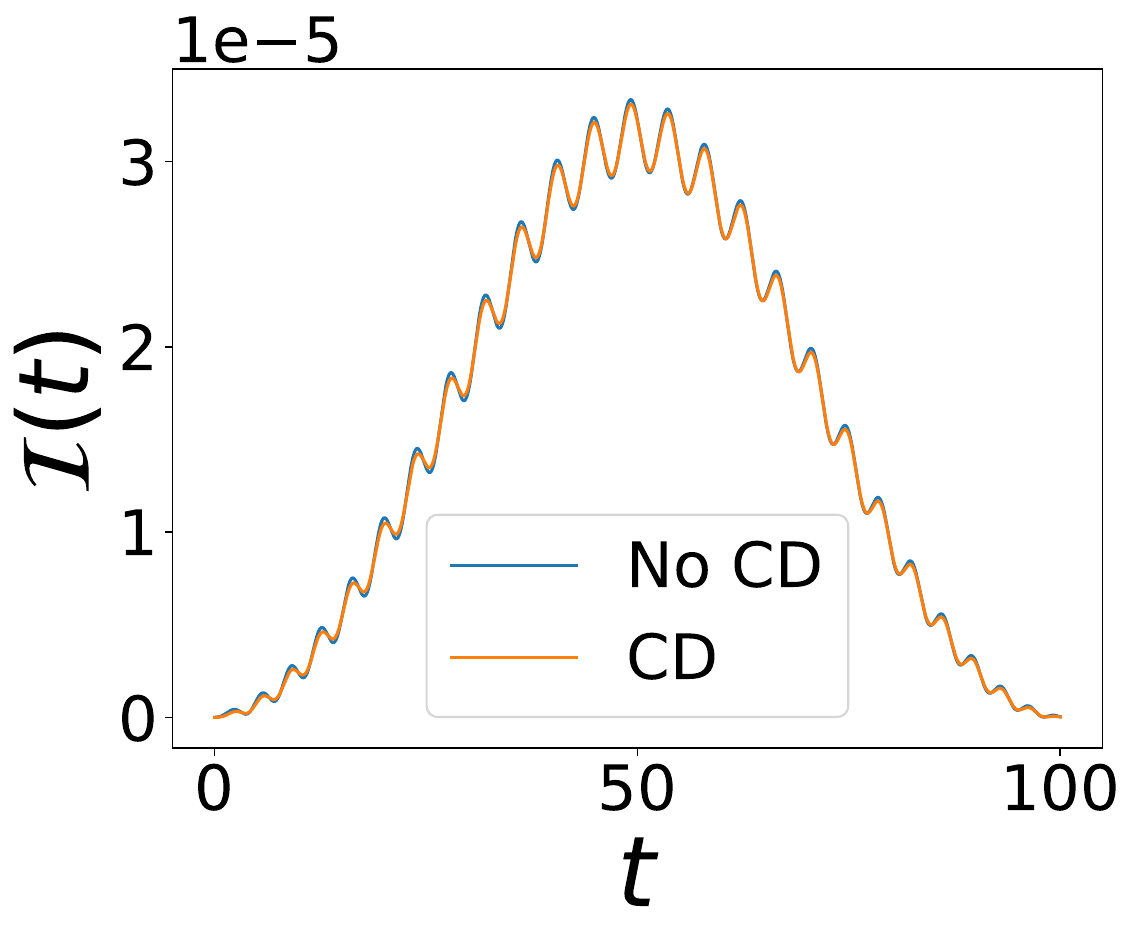}
    \caption{(a, b) Infidelity versus evolution time $T$ for $\Delta t=0.1$ (purple line), $\Delta t= 0.01$ (yellow line) and $\Delta t= 0.001$ (blue line), for a single qubit. In panel (a), there are no counterdiabatic corrections. In panel (b), we include the exact CD potential. The axes are in logarithmic scale. By comparing these two panels, we highlight the difference between the fit $T^{-\beta}$ proposed in Ref.~\cite{Kovalsky2023} (panel~(a)), corresponding to evolutions without the CD potential, and the fit introduced in Eq.~\eqref{eq:perturbation-theory-T} (panel~(b)), which applies to evolutions with the CD potential. We observe that our proposed fit accurately captures the error at shorter evolution times compared to the former case.
    (c, d) Dynamics of the infidelity, $\Delta t=0.01$. In these plots, we show the dynamics for a single qubit. Blue curves represent the dynamics without CD corrections, orange curves represent the dynamics with the exact CD potential. In panel (c), $T=1$; in panel (d), $T=100$. Panel~(d) illustrates the concept of self-healing behavior introduced in Ref.~\cite{Kovalsky2023} valid in the limit of $T\rightarrow\infty$: the error initially increases, reaches a maximum, and subsequently decreases, eventually becoming comparable to its initial value at the end of the evolution. For $T = 100$, this behavior is observed for both dynamics, with and without the CD potential, as both regimes are effectively adiabatic. In contrast, for shorter evolution times, only the dynamics with the CD potential remain adiabatic. Accordingly, as shown in panel~(c), the self-healing behavior of the error emerges exclusively in this case.}
    \label{fig:non-interacting model-infidelity-dynamics-non-interacting model}
\end{figure}

\section{Results}\label{sec:Examples}

\subsection{Noninteracting model}\label{sec:noninteracting}

We first benchmark our analysis on a noninteracting system, namely a single qubit. For this model, the CD correction generates an additional rotation around the $y$ axis of the Bloch sphere (i.e., the exact AGP is proportional to $\sigma^{y}$), providing a clean setting to isolate digitization effects.

In the upper row of Fig.~\ref{fig:non-interacting model-infidelity-dynamics-non-interacting model}, we report the final groundstate infidelity $\mathcal{I}(T)$ [Eq.~\eqref{eq:digital-infidelity}] as a function of the total runtime $T$. Panel (a) shows $\mathcal{I}(T)$ for digitized adiabatic evolution without CD, while panel (b) shows the corresponding CD-assisted digitized evolution. The blue, yellow, and violet markers correspond to $\Delta t = 10^{-3}$, $10^{-2}$, and $10^{-1}$, respectively. As expected for first-order product formulas, the infidelity increases with $\Delta t$, consistent with an overall $\mathcal{O}(\Delta t^2)$ dependence.

In the large-$T$ regime, the curves with and without CD approach each other. This is expected because $\dot{\lambda}\sim 1/T$, so the CD contribution is parametrically small for slow ramps and both protocols closely follow the same adiabatic path. At shorter runtimes, however, including the CD term systematically reduces the infidelity for all $\Delta t$. This reduction is nontrivial: although CD adds an extra generator to the digitized sequence, it suppresses diabatic transitions and thereby makes the residual error more purely discretization-driven, enhancing the phase-cancellation mechanism underlying self-healing.

To assess self-healing in its original form, we first focus on the evolution without CD [panel (a)]. In the long-time regime, the self-healing prediction implies a scaling $\mathcal{I}(T)\propto T^{-2}\Delta t^2$ for fixed $\Delta t$~\cite{Kovalsky2023}. Consistently, a power-law fit $\mathcal{I}(T)\propto T^{\beta}$ over the largest-$T$ data (here $T>10$) yields $\beta\simeq-2$ (within fit uncertainty), in agreement with the asymptotic self-healing behavior. For smaller $T$, deviations from a pure power law become visible, indicating the onset of finite-time effects beyond the asymptotic regime.

In panel (b), we test the finite-time theory by fitting the CD-assisted data to the analytical expression in Eq.~\eqref{eq:perturbation-theory-T}. The fit parameters are the effective gap $\Delta_{10}$ and the dominant Fourier index $\bar{q}$. The rationale is that, if Eq.~\eqref{eq:perturbation-theory-T} captures the leading contribution, the extracted parameters should be compatible with the characteristic gap scale and the leading harmonic in the effective perturbation. We find that the fit describes the data well for $T\gtrsim 1$, yielding $\bar{q}_{\rm fit}\approx 1$ and $\Delta_{10,{\rm fit}}\approx 1.16,\,1.14,$ and $1.12$ for $\Delta t=0.1,\,0.01,$ and $0.001$, respectively. While the fitted gap is not expected to coincide exactly with the bare instantaneous gap (since the expression is derived under controlled approximations), its magnitude is consistent with the relevant excitation scale of the single-qubit spectrum, supporting the validity of the harmonic-perturbation description in this regime.

Figures~\ref{fig:non-interacting model-infidelity-dynamics-non-interacting model}(c) and (d) show the time-resolved digitization error $\mathcal{I}(t)$, see Eq. \eqref{eq:digital-infidelity}, for a fixed step $\Delta t=0.01$, comparing the adiabatic digitized protocol (blue) and the CD-assisted digitized protocol (orange). For $T=100$ [panel (d)], both evolutions lie deep in the adiabatic regime and exhibit similarly small infidelity throughout the protocol. For $T=1$ [panel (c)], the difference is pronounced: without CD the infidelity grows during the ramp and leads to a larger final error, whereas the CD-assisted evolution yields a substantially smaller final infidelity. Importantly, the CD-assisted curve displays a bounded, oscillatory-in-time structure, consistent with an error dominated by the discretization-induced mixing term $\mathcal{R}_{ij}$, while the uncorrected evolution receives additional finite-$T$ contributions (encoded by $\mathcal{S}_{ij}$ and $\mathcal{Q}_{ij}$) that lead to a more cumulative error growth~\cite{Kovalsky2023}.

Therefore, the single-qubit benchmark provides a transparent demonstration of finite-time self-healing: suppressing diabatic transitions with CD driving isolates the digitization-induced contribution and reveals the bounded, interference-driven structure underlying self-healing at finite $T$. Having established the mechanism in the noninteracting limit, we now turn to interacting spin models to examine its robustness across qualitatively different dynamical regimes.

\subsection{Interacting models}

In this section we turn to interacting models, focusing on the fully connected Ising model obtained from Eq.~\eqref{eq:hp} at $p=2$ and the Ising chain in Eq.~\eqref{eq:hz}. In both cases, the target Hamiltonians contain two-body interactions. For the Ising chain we implement the exact CD potential [Eq.~\eqref{eq:CD-real-potential}], whereas for the fully connected model we use the variational CD construction described in Sec.~\ref{sec:Counterdiabatic-potential}. As in the noninteracting benchmark, we test the analytical picture developed in Sec.~\ref{sec:analytical-description} and show that, once diabatic transitions are suppressed, the residual digitization error exhibits a bounded oscillatory (harmonic) structure.

Here, we compare the analytical prediction in Eq.~\eqref{eq:perturbation-theory-T} with numerical simulations. Before doing so, we stress that Eq.~\eqref{eq:perturbation-theory-T} is an approximate description for two reasons. First, the connection between the infidelity and instantaneous transition amplitudes is, in general, only an inequality: as implied by Eq.~\eqref{eq:dis-infidelity}, the final infidelity receives contributions from multiple excited levels, and even when all levels are included the relation remains a bound (it becomes an equality only in the single-qubit case). Second, the step from the harmonic expansion~\eqref{eq:harmonic-potential} to Eq.~\eqref{eq:perturbation-theory-T} assumes that the residual mixing can be treated perturbatively and that a single Fourier component dominates. When several harmonics contribute with comparable weight, Eq.~\eqref{eq:perturbation-theory-T} should be interpreted as describing an effective dominant mode that captures the leading scaling behavior rather than a unique microscopic frequency component.

\begin{figure}
    \hspace{\textwidth}(a)\hspace{0.23\textwidth}(b)
    \includegraphics[width=0.23\textwidth]{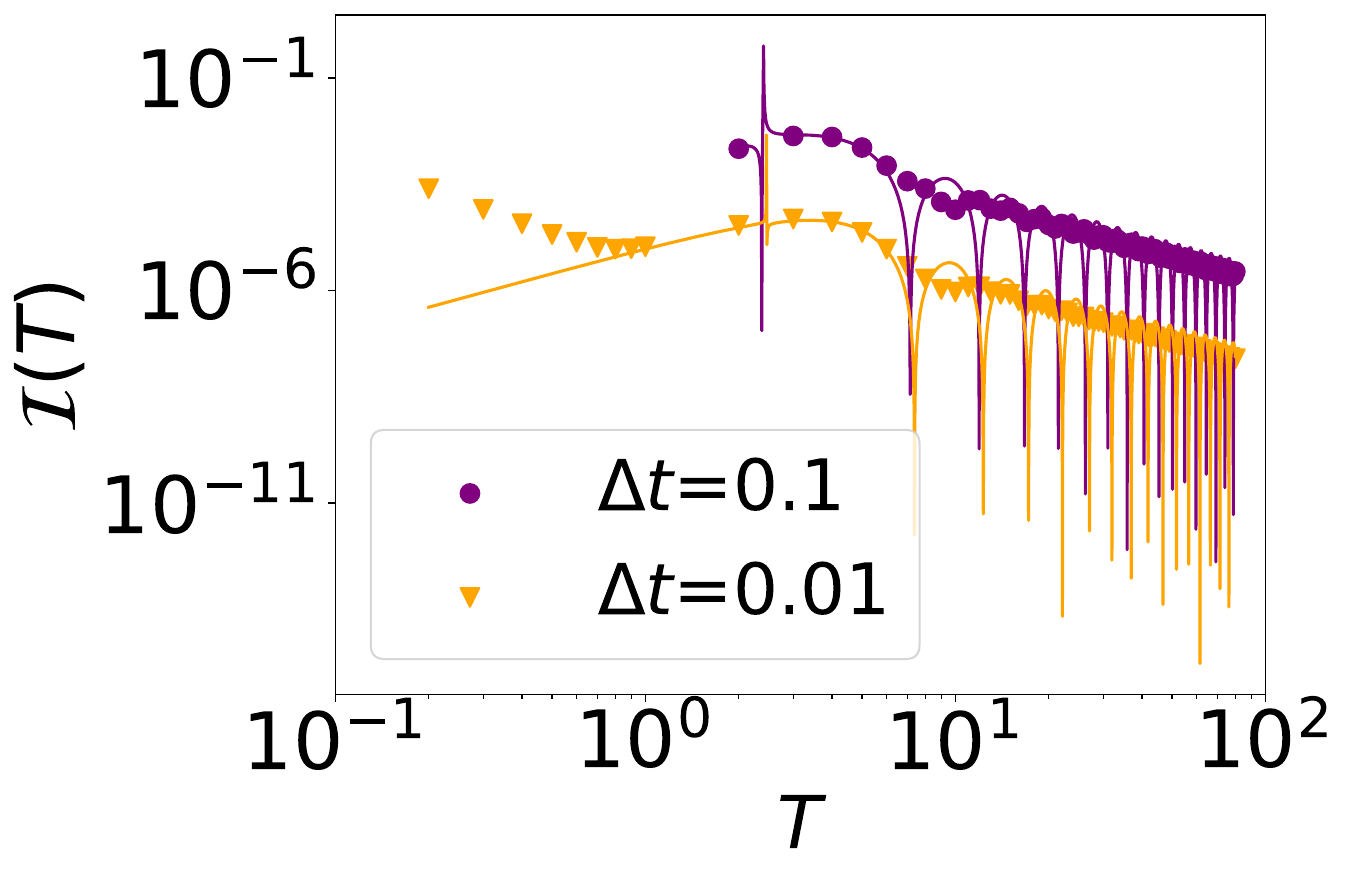}
    \includegraphics[width=0.23\textwidth]{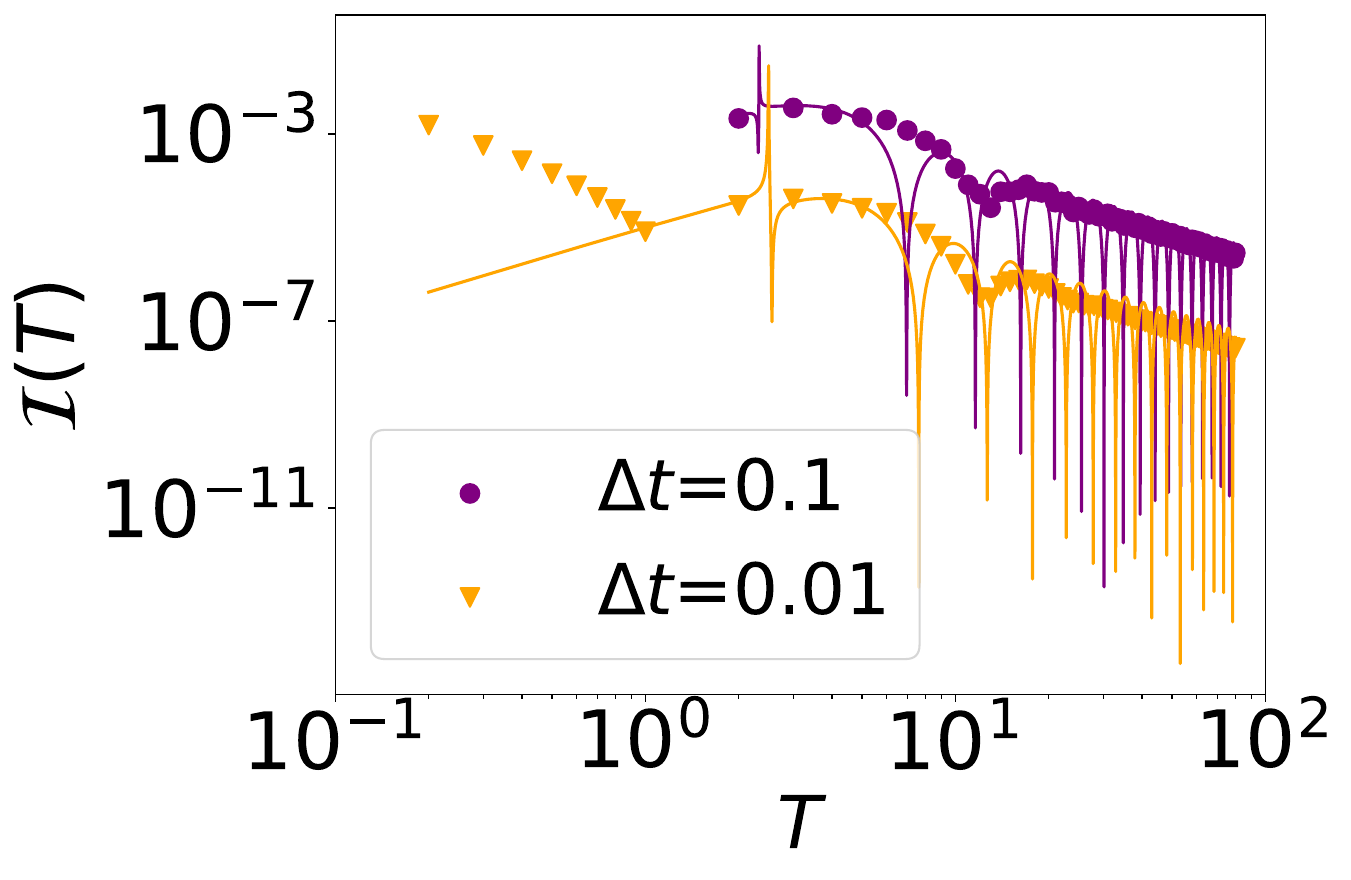}
    (c)\hspace{0.25\textwidth}(d)\hspace{0.25\textwidth}
    \includegraphics[width=0.23\textwidth]{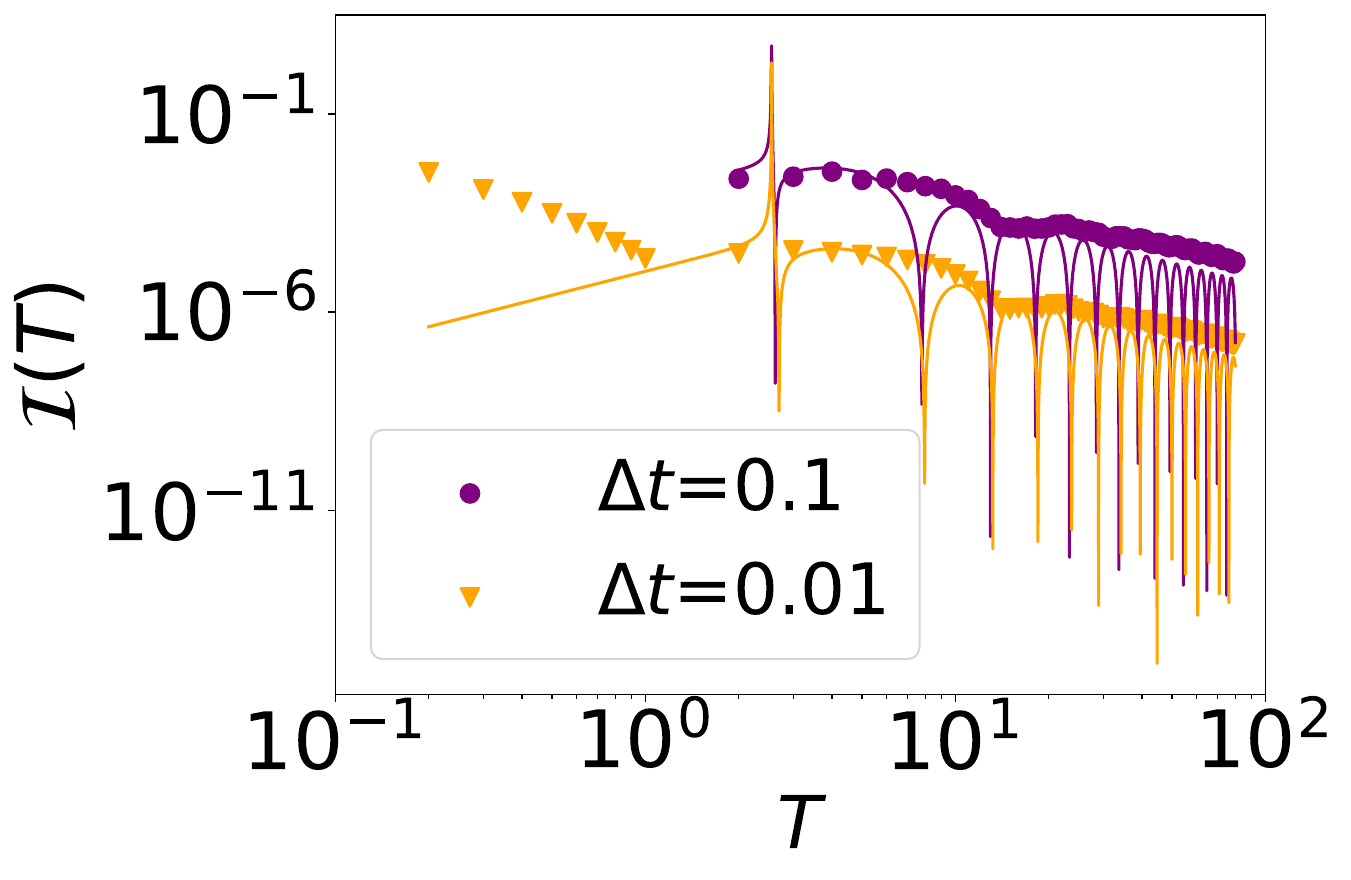}
    \includegraphics[width=0.23\textwidth]{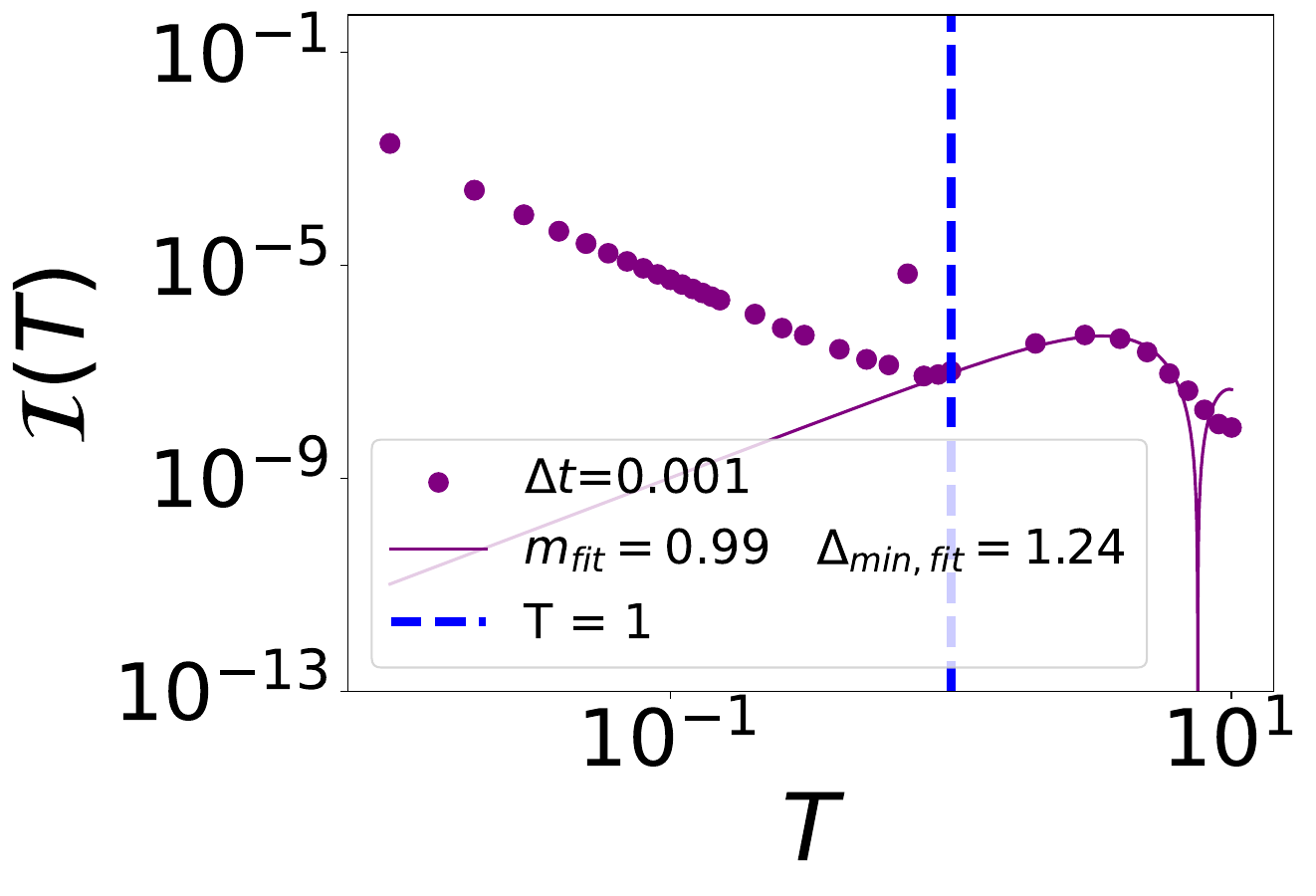}
    \caption{Infidelity versus final time $T$ for the Ising model with the exact CD potential. Orange triangles represent data for $\Delta t = 0.01$, while purple circles correspond to $\Delta t=0.1$. Continuous lines represent numerical fits done using the analytical prediction given by Eq.~\eqref{eq:perturbation-theory-T}. (a) $J_Z = 0.1$  (b) $J_Z = 0.5$ (c) $J_Z=1$. (d) Infidelity vs $T$ for $\Delta t=0.001$ for $J_Z = 0.1$. 
    In these plots we consider a system size $N=6$. Panels~(a)--(c) demonstrate that the fit predicted in Eq.~\eqref{eq:perturbation-theory-T} remains valid also for interacting models. Panel~(d), instead, highlights the limitation of the theory, showing that it breaks down for $T < 1$.}
    \label{fig:fit-non-intercting-spin-6}
\end{figure}

\subsubsection{Ising model}

In Figs.~\ref{fig:fit-non-intercting-spin-6}(a-c), we plot the final infidelity $\mathcal{I}(T)$ as a function of the total runtime $T$ for the Ising chain at $N=6$, using the digitized evolution with the 
exact CD potential. The orange and violet datasets correspond to Trotter steps $\Delta t=0.01$ and $\Delta t=0.1$, respectively. Panels (a), (b), and (c) show $J_Z=0.1$, $0.5$, and $1$. Symbols denote numerical data, while solid curves show fits to Eq.~\eqref{eq:perturbation-theory-T}, treating the dominant harmonic index $\bar{q}$ and the effective gap scale $\Delta_{10}$ as fitting parameters. We emphasize that Eq.~\eqref{eq:perturbation-theory-T} is used here as a leading-order effective description: because the infidelity generally receives contributions from multiple excited levels and, in addition, several harmonics may contribute to the residual mixing, quantitative agreement is not expected to be exact. Nevertheless, the fits provide a stringent test of whether the predicted oscillatory structure and scaling with $T$ and $\Delta t$ are captured by the data.

To assess consistency between the analytical picture and the numerical results, we analyze the fitted parameters $(\bar{q}_{\rm fit},\Delta_{10,{\rm fit}})$. Tables~\ref{tab:fit parameters-delta-0.1} and~\ref{tab:fit parameters-delta-0.01} summarize these values for $J_Z=0.1,0.5,1$ and system sizes $N=3,4,5,6$ at $\Delta t=0.1$ and $\Delta t=0.01$, respectively (here we highlight $N=6$; the remaining cases are reported in the Supplementary Material). The parameter $\bar{q}_{\rm fit}$ corresponds to the dominant Fourier index in Eq.~\eqref{eq:perturbation-theory-T}, while $\Delta_{10,{\rm fit}}$ provides an effective gap scale entering the phase factor and the denominator of the perturbative expression.

\begin{figure}
    \hspace{\textwidth}(a)\hspace{0.32\textwidth}(b)\hfill\\
    \centering
    \includegraphics[width=0.49\linewidth]{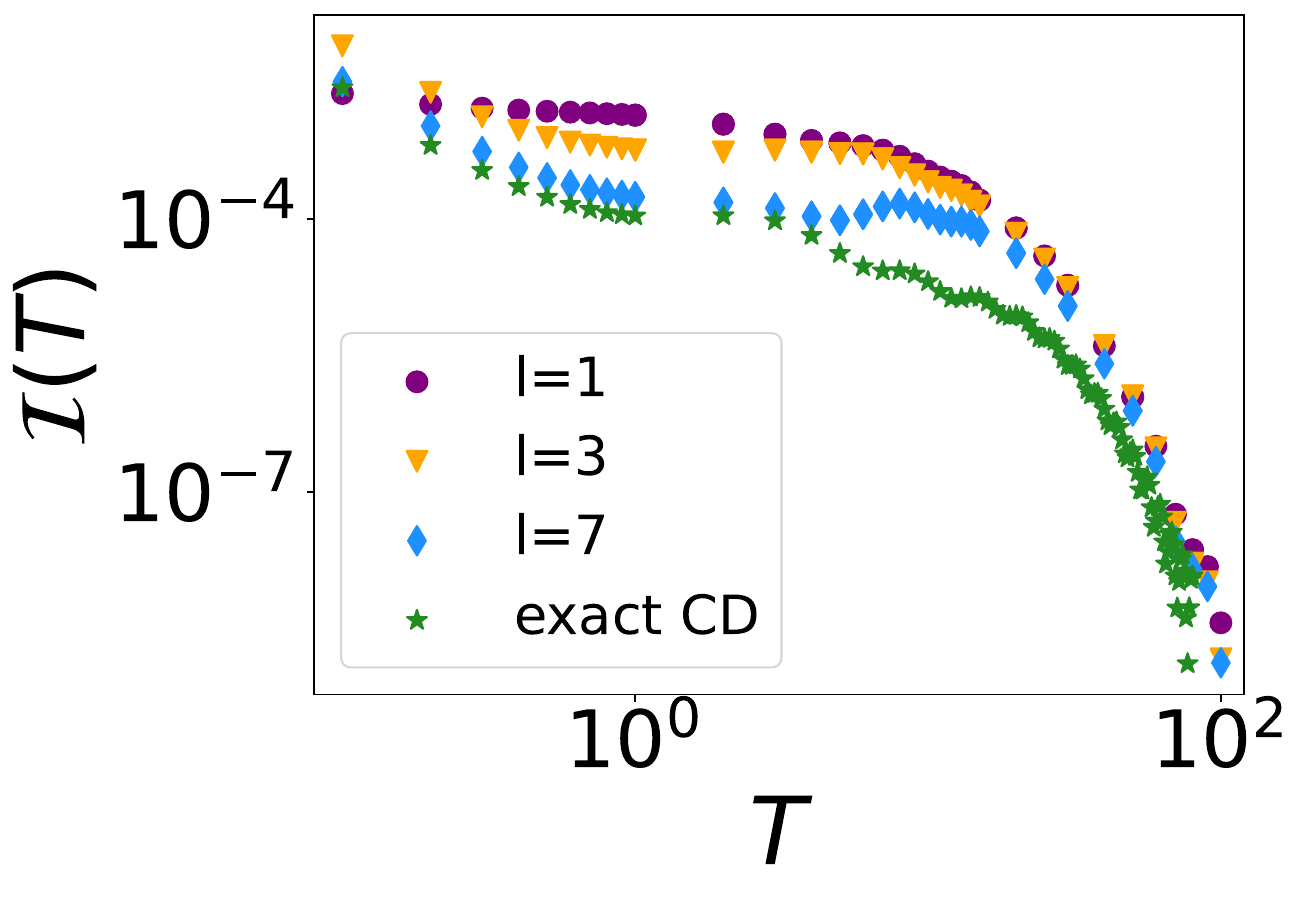}
    \includegraphics[width=0.49\linewidth]{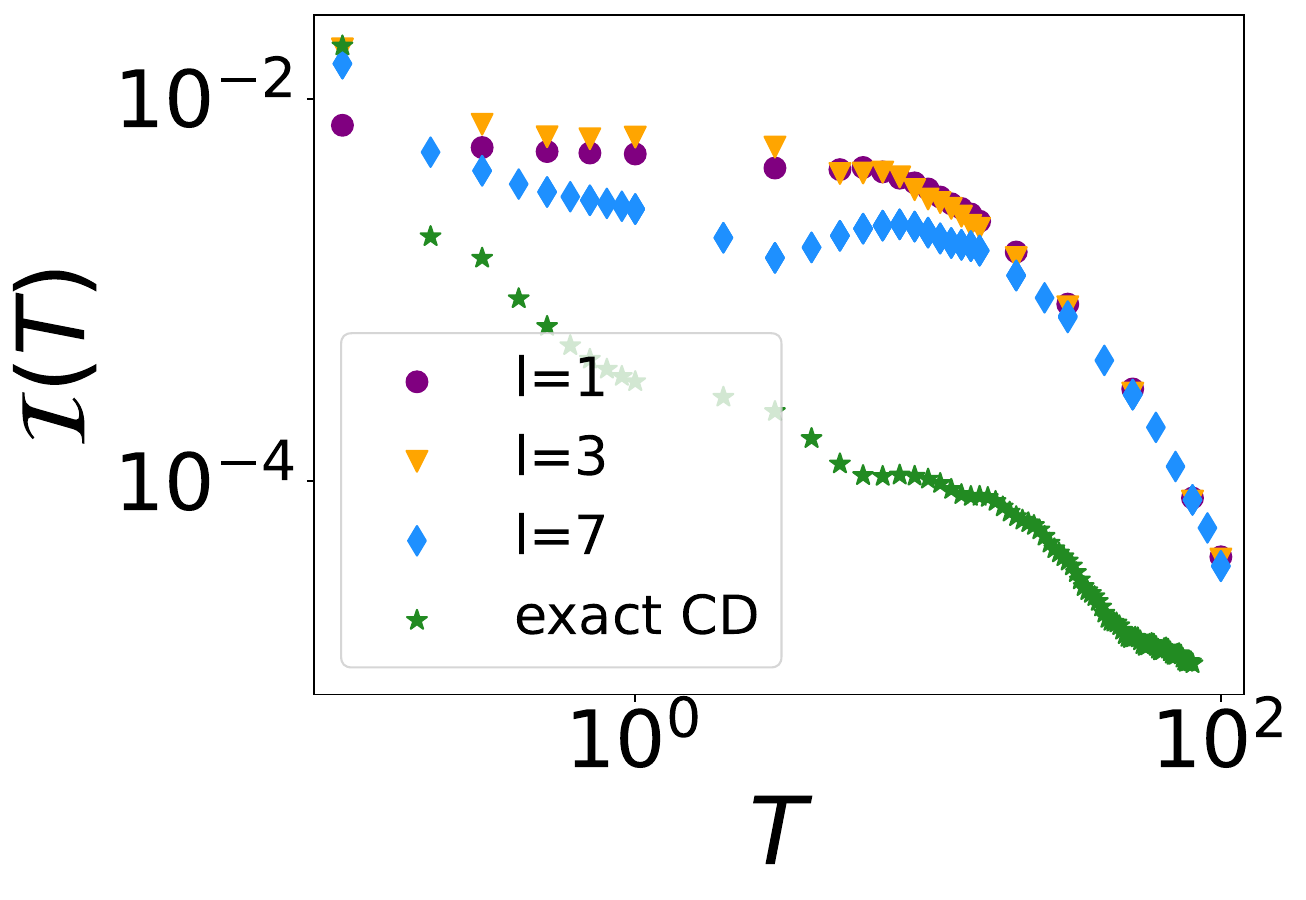}
    \caption{
    Final infidelity $\mathcal{I}(T)$ for the $p$-spin ($p=2$) model with $\Delta t = 0.01$. (a) $N = 10$; (b) $N = 30$. The green points denote the evolution generated by the exact counterdiabatic (CD) potential in Eq.~\eqref{eq:CD-real-potential}, while the other points correspond to the evolution defined in Eq.~\eqref{eq:CD-digitalized-evolution}, obtained using a variational CD potential. As expected, higher-order CD corrections become increasingly significant at smaller values of $T$, with the highest-order approximation providing the closest agreement with the exact solution.
    }
    \label{fig:p-spin-model}
\end{figure}
\begin{table}[h]
 \caption{Pairs $(\bar{q}_\text{fit},\Delta_\text{min,fit})$ for Ising systems with sizes $N=3,4,5,6$ and couplings $J_Z=0.1,0.5,1$. In this table we fix $\Delta t=0.1$.}
\setlength{\tabcolsep}{3.8pt}
    \centering
    \begin{tabular}{cccc}
    \toprule
    & $J_Z=0.1$ & $J_Z=0.5$ & $J_Z=1$\\ 
    \midrule
    $N=3$ & $(0.9976,1.5880)$ & $(0.9865,1.6916)$ & $(0.9750,1.7579)$ \\
    $N=4$ & $(1.0068,1.4828)$ & $(1.0000,1.5707)$ & $(0.9963,1.6284)$\\
    $N=5$ & $(1.0072,1.3891)$ & $(1.0048,1.4678)$ & $(1.0011,1.5272)$ \\
    $N=6$ & $(1.0060,1.3151)$ & $(1.0018,1.3535)$ & $(0.9860,1.2139)$\\
    \bottomrule
    \end{tabular}
    \label{tab:fit parameters-delta-0.1}
\end{table}

\begin{figure*}
    \hspace{\textwidth}(a)\hspace{0.32\textwidth}(b)\hspace{0.32\textwidth}(c)\hfill\\
    \centering
    \includegraphics[width=0.34\textwidth]{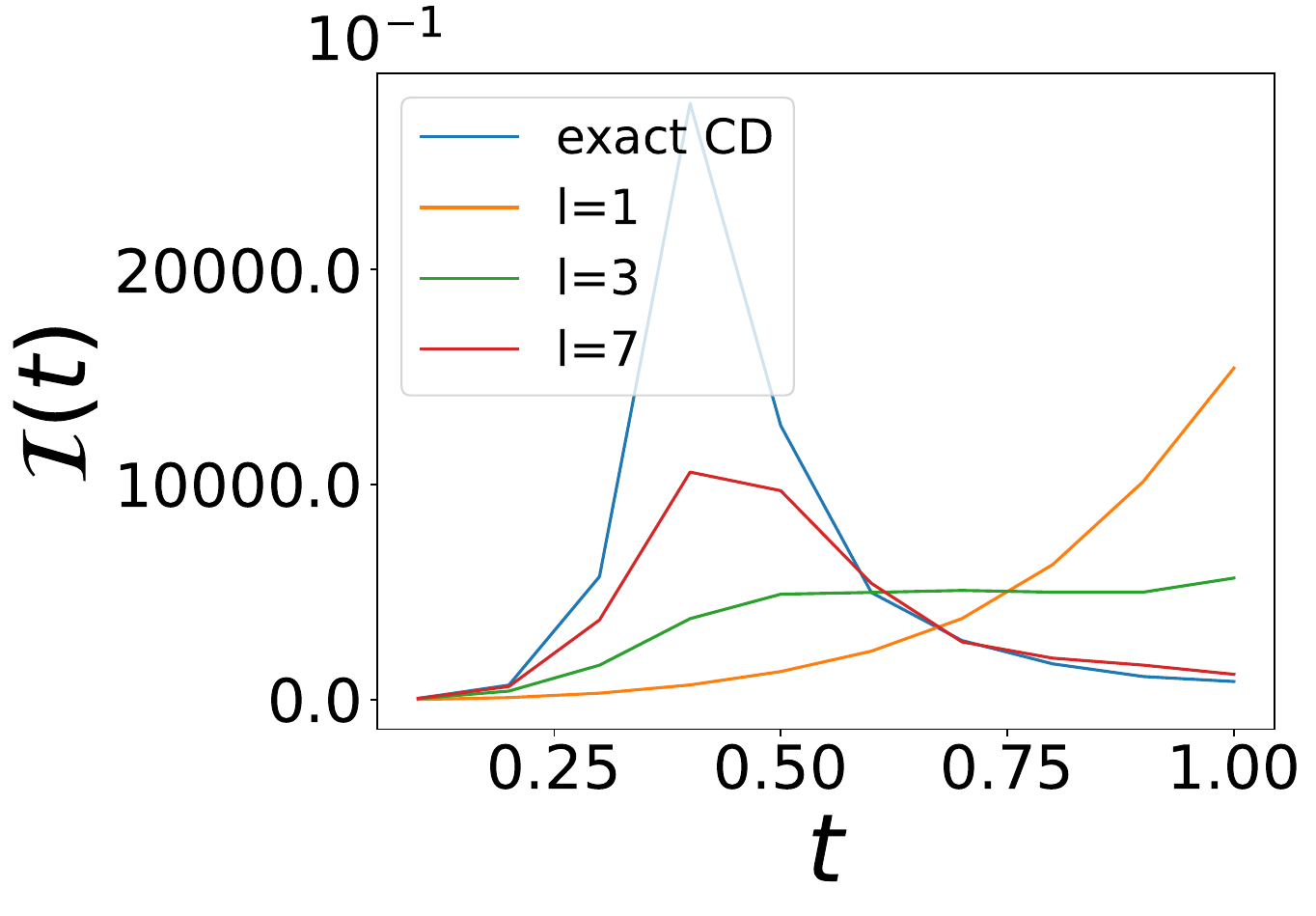}
    \includegraphics[width=0.32\textwidth]{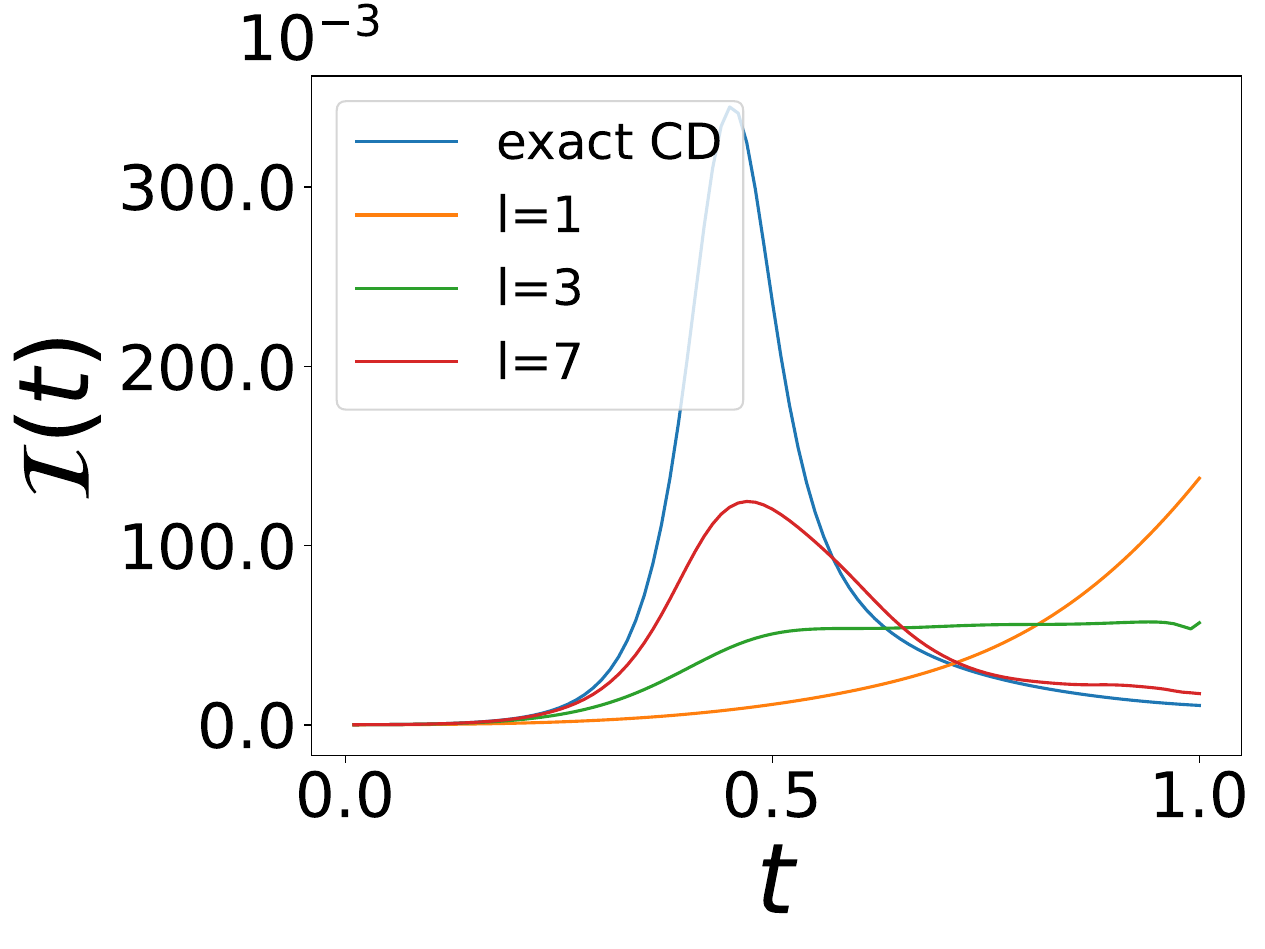}
    \includegraphics[width=0.3\textwidth]{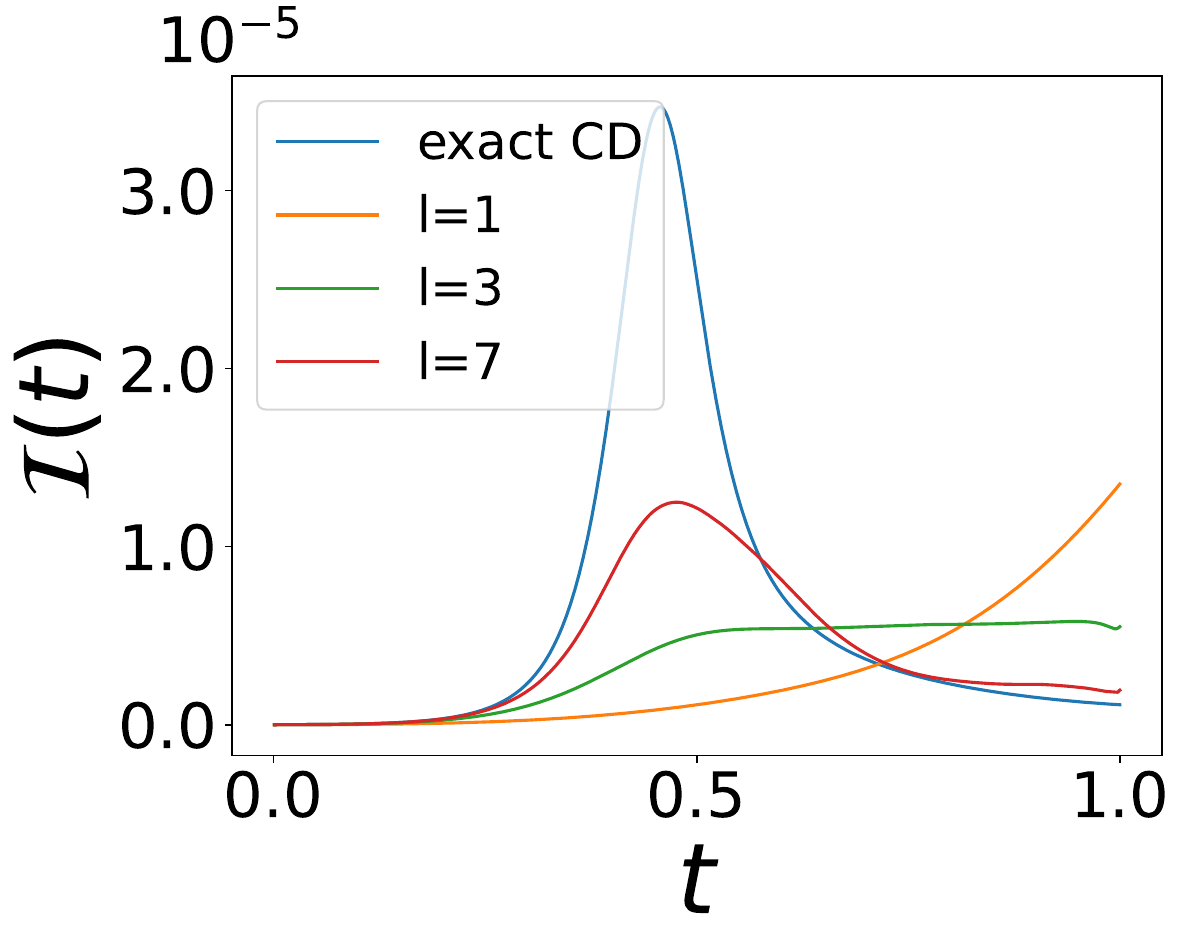}
    \caption{Infidelity $\mathcal{I}(t)$ versus time for $t\in\left[0,T\right]$, for $T=1$. The evolution is for a $p$-spin model for $p=2$ and size $N=10$. The blue curves represent the evolution generated by the exact counterdiabatic (CD) potential in Eq.~\eqref{eq:CD-real-potential}, while the other curves correspond to the evolution defined in Eq.~\eqref{eq:CD-digitalized-evolution} obtained using a variational CD potential. Different colors denote different truncation orders $l$ in the nested commutator expansion: orange for $l=1$, green for $l=3$, and red for $l=7$. (c) $\Delta t= 0.001$. We see that, as the order of approximation increases, for $t\rightarrow T$, $\mathcal{I}(T)$ approaches zero for all $\Delta t$ and the error starts to have a self-healing behavior, analogous to the case of the exact CD potential.}
    \label{fig:p-2-spin-model-dynamics}
\end{figure*}

Since the approximations discussed above are in place, both $\bar{q}_{\rm fit}$ and $\Delta_{\rm min,fit}$ should be interpreted as {effective} parameters rather than microscopic quantities. Specifically, $\Delta_{\rm min,fit}$ captures the net contribution of multiple excitation channels (i.e., the set of gaps between the groundstate and the relevant excited levels that enter the infidelity), while $\bar{q}_{\rm fit}$ effectively summarizes the dominant Fourier content of the residual mixing in Eq.~\eqref{eq:perturbation-theory-T}. When several harmonics contribute, $\bar{q}_{\rm fit}$ should be viewed as identifying an effective leading mode rather than a unique microscopic frequency.

Despite this simplification, we find that $\bar{q}_{\rm fit}$ clusters around $1$ across all parameters considered, consistent with the expectation that the leading harmonic index is an integer and indicating that the lowest Fourier component dominates the residual error in the fitted regime. For $\Delta_{\rm min,fit}$, we obtain values that remain of the same order as the characteristic excitation gap scale of the model, supporting the interpretation of Eq.~\eqref{eq:perturbation-theory-T} as a controlled leading-order description. Finally, the fit quality deteriorates at sufficiently small $T$, where the perturbative assumptions underlying Eq.~\eqref{eq:perturbation-theory-T} are no longer satisfied.

Figure~\ref{fig:fit-non-intercting-spin-6}(d) provides a representative illustration of this breakdown for $J_Z=0.1$ (exact CD). In particular, it highlights the onset of deviations around $T\sim 1$, beyond which the simple perturbative expression ceases to capture the numerical data quantitatively. This is consistent with the analysis in Sec.~\ref{sec:analytical-description}: for short runtimes, higher-order contributions (including terms scaling as $\mathcal{O}(1/T^2)$ in the adiabatic-frame expansion) become non-negligible, and additional Fourier components can contribute appreciably. Consequently, deviations from Eq.~\eqref{eq:perturbation-theory-T} at $T\lesssim 1$ should be interpreted as the breakdown of the leading-harmonic perturbative regime and the onset of genuinely nonperturbative finite-time dynamics.

%

\begin{table}[t]
\caption{Pairs $(\bar{q}_\text{fit},\Delta_\text{min,fit})$ for Ising systems with sizes $N=3,4,5,6$ and couplings $J_Z=0.1,0.5,1$. In this table we fix $\Delta t=0.01$.}
\setlength{\tabcolsep}{3.8pt}
    \centering
    \begin{tabular}{cccc}
    \toprule
    & $J_Z=0.1$ & $J_Z=0.5$ & $J_Z=1$ \\
    \midrule
    $N=3$ & $(1.0004,1.5635)$ & $(0.8578,1.5803)$ & $(0.7191,1.5259)$ \\
    $N=4$ & $(1.0020,1.4374)$ & $(1.0013,1.5115)$ & $(0.9977,1.4695)$ \\
    $N=5$ & $(0.9987,1.33.96)$ & $(0.9907,1.3160)$ & $(0.9707,1.2094)$ \\
    $N=6$ & $(0.9993,1.2767)$ & $(0.9875,1.2428)$ & $(0.9725,1.1965)$ \\
    \bottomrule
    \end{tabular}
    \label{tab:fit parameters-delta-0.01}
\end{table}
\begin{figure}[h]
    \centering
    \includegraphics[width=0.7\linewidth]{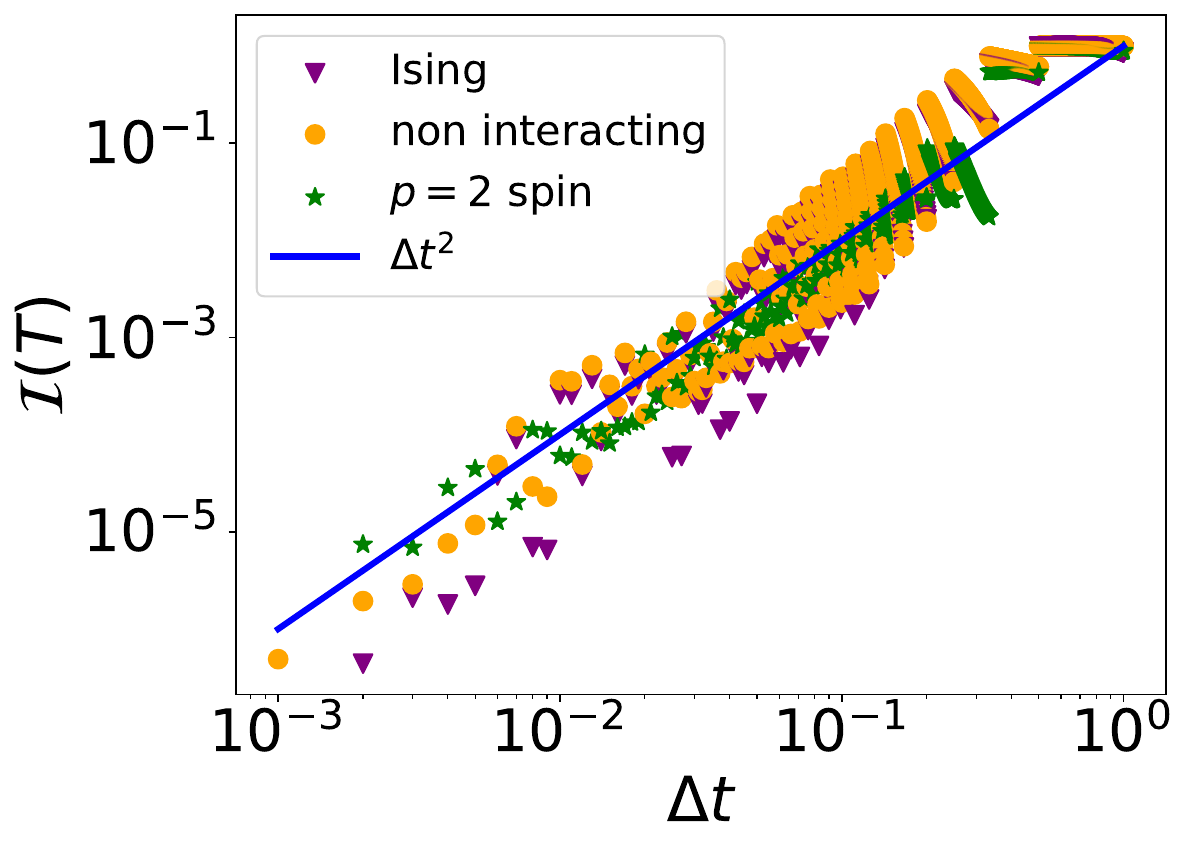}
    \caption{Infidelity $\mathcal{I}(T)$ as a function of the time step $\Delta t$. Violet points correspond to the Ising model with $N=6$ and $J_Z=0.1$. Yellow points represent the noninteracting case. Green points refer to the interacting $p$-spin model with $p=2$. The blue line indicates the expected $\Delta t^2$ scaling of the Trotter error. Both the evolution and the Trotterization are performed with the inclusion of the CD potential. The total evolution time is fixed at $T=1$. The axes are in logarithmic scale. This plot confirms the quadratic dependence of the error on $\Delta t$, i.e., $\propto \Delta t^2$. On the other hand, in this work we focus on the dependence of the error on $T$ and on how it can be reduced using Shortcuts to Adiabaticity techniques.}
    \label{fig:delta-vs-infidelity}
\end{figure}

\subsubsection{$p$-spin model}

We now focus on the $p$-spin model at $p=2$ (i.e., the fully connected Ising model). Fig.~\ref{fig:p-spin-model} reports the final-time infidelity $\mathcal{I}(T)$ obtained with the {variational} CD potential for different truncation orders: violet, yellow, and blue markers correspond to $l=1$, $l=3$, and $l=7$, respectively. Panels (a) and (b) show system sizes $N=10$ and $N=30$. The Trotter step is fixed to $\Delta t=0.01$. As expected, the impact of the CD correction is most pronounced at shorter evolution times: increasing $l$ systematically lowers the final infidelity for $T$ in the nonadiabatic/finite-time regime, while for large $T$ the curves for different $l$ approach each other as the dynamics becomes increasingly adiabatic and the CD contribution $\propto \dot{\lambda}$ becomes less relevant. Although the approximate CD term prevents a quantitative comparison with the single-mode expression~\eqref{eq:perturbation-theory-T}, the data reproduce the same qualitative trends as in the exact-CD case, with higher truncation orders yielding progressively stronger suppression of finite-time errors.

To further probe the regime where the variational CD term has the largest dynamical impact, we fix $T=1$ and analyze the time-resolved infidelity. Fig.~\ref{fig:p-2-spin-model-dynamics} shows $\mathcal{I}(t)$ for $p=2$ and $N=10$ under the CD-assisted digitized evolution~\eqref{eq:CD-digitalized-evolution}, using the variational CD potential described in Sec.~\ref{sec:Counterdiabatic-potential}. From left to right the Trotter step decreases from $\Delta t=0.1$ to $0.01$ and $0.001$. While the overall magnitude of the error decreases with $\Delta t$ due to reduced discretization, the qualitative time dependence is similar across panels. Increasing the truncation order improves the suppression of diabatic transitions and drives the infidelity closer to zero near the end of the protocol. In particular, at $l=7$ the infidelity exhibits a bounded, nonmonotonic profile---an initial rise followed by a decay toward $t\to T$, indicating that the CD-assisted protocol converts the dominant error into a largely discretization-driven contribution whose net effect is partially canceled by interference at the end of the evolution.

Finally, we summarize the $\Delta t$-dependence of the final error across the different models. As illustrated in Fig.~\ref{fig:delta-vs-infidelity}, the final-time infidelity increases with $\Delta t$ and follows a clear quadratic scaling, $\mathcal{I}(T) \propto \Delta t^2$, for the noninteracting model, the Ising chain (e.g., $N=6$, $J_Z=0.1$), and the $p=2$ fully connected model (e.g., $N=10$). This confirms that the leading discretization contribution remains robust within our framework, while the role of (exact or approximate) CD driving is to suppress diabatic errors and thereby expose the intrinsic, interference-driven structure underlying finite-time self-healing.

\section{Discussion and Conclusion}\label{sec:conclusions}

In this work we investigated product-formula (Trotter) errors in digitized simulations of time-dependent (gapped) Hamiltonians, with a focus on adiabatic state preparation at {finite} runtime. Beyond the long-time asymptotic regime where self-healing was previously established, we showed that a self-healing-like suppression of discretization errors persists at finite $T$ once diabatic contributions are properly controlled. Using both noninteracting and interacting spin models, and combining analytical arguments with numerical benchmarks, we characterized how the residual error depends jointly on the Trotter step $\Delta t$ and the total runtime $T$.

A key conceptual message is that CD driving acts not only as a control resource but also as a {diagnostic knob} for digitized dynamics. By suppressing the nonadiabatic channels that scale as $\mathcal{O}(1/T)$ (encoded by the $\mathcal{S}_{ij}$ and $\mathcal{Q}_{ij}$ terms in the instantaneous-eigenbasis description), CD isolates the discretization-induced mixing channel governed by $\mathcal{R}_{ij}$. In this regime, the remaining error budget is predominantly digital: it is set by the product-formula structure and scales with $\Delta t$, enabling a clean separation between dynamical (finite-$T$) and digitization (finite-$\Delta t$) contributions. This separation provides a transparent framework for interpreting digitized adiabatic protocols and for identifying the microscopic origin of finite-time self-healing.

Our analysis further clarifies in what sense finite-time self-healing is generic. The mechanism relies on two physically transparent ingredients: (i) once the diabatic channels are suppressed, the residual digitization-induced mixing can be treated as a weak, smooth function along the schedule parameter, and therefore admits an effective harmonic (Fourier) representation; and (ii) the accumulated effect of this mixing is dominated by interference, such that successive contributions can partially cancel at the end of the protocol rather than accumulate monotonically. When endpoint-smooth schedules are used (e.g., $\dot{\lambda}(0)=\dot{\lambda}(T)=0$), the residual mixing naturally satisfies boundary constraints and a sine-series representation becomes particularly natural. Away from near-resonant windows in which $|\Delta_{i0}\pm \pi q/T|$ becomes small, the resulting interference yields a bounded, oscillatory structure in time and the characteristic envelope $\sim T^{-2}\Delta t^{2}$ for first-order digitization. These criteria provide a set of testable conditions---rather than a qualitative slogan---for when finite-time self-healing should be expected on a given platform.

An important physical implication is the revised role of the energy gap under CD-assisted digitization. In conventional adiabatic evolution, the gap primarily limits performance through diabatic transitions. Under accurate CD driving, diabatic channels are suppressed and the gap enters mainly through {phase accumulation} and {near-resonance structure}. In the instantaneous eigenbasis of the reference Hamiltonian $H(\lambda)$, phases of the form
$\varphi_{i0}=\int d\lambda\,\Delta_{i0}(\lambda)/\dot{\lambda}$
govern whether contributions interfere destructively (self-healing) or constructively, while the analytic expressions expose resonance-like enhancement when $\Delta_{i0}\approx \pi q/T$. In this sense, CD effectively converts the gap from a parameter that controls adiabaticity into one that shapes an {interference landscape} for digital errors at finite runtime.

Our interacting-model results also address scalability and the role of approximate CD constructions. While exact CD is generally unavailable in many-body settings, variational CD can substantially suppress diabatic channels and drive the dynamics into an interference-dominated regime where the residual error is primarily discretization controlled. Increasing the truncation order improves this suppression and enhances the approach to the bounded, oscillatory behavior predicted by the effective harmonic description. This observation is particularly relevant near critical regions, where the exact adiabatic gauge potential becomes increasingly nonlocal and low-order ans\"atze may be insufficient, emphasizing a practical crossover from transition-dominated error accumulation to digitization-dominated self-healing.

In general terms, this paper demonstrates that it is possible to implement time-dependent dynamics on gate-based quantum computers, including short-time evolutions, with a controllable Trotter error. Previously, it was widely believed that the digitization of a time-dependent Hamiltonian leads to an accumulation of error: as the evolution time $t$ increases, the growing number of unitary operations causes the simulated evolution to deviate further from the exact one. In particular, the norm distance $||U(t) - U_{\mathrm{T}}(t)||$ between the exact time-evolution operator (see Eq.~\eqref{eq:real-evolution-operator}), obtained from the Schr\"odinger equation, and its Trotterized counterpart (see Eq.~\eqref{eq:digitalized-evolution}) was expected to increase with $t$. However, Ref.~\cite{Kovalsky2023} shows that the error decreases for long-time evolutions, demonstrating that digitized time-dependent dynamics do not necessarily accumulate errors over time. Nevertheless, implementations in long-time evolutions require a large number of Trotter steps, which worsens issues related to decoherence and noise. 

In this paper, we generalize the results of Ref.~\cite{Kovalsky2023} to both short and long-time evolutions by considering dynamics driven by CD terms. When the goal of a quantum circuit is to reach a target final state, CD driving modifies the evolution path while preserving the final state. In other words, we showed that the self-healing property is closely related to the adiabaticity of the evolution, and that CD driving enforces adiabatic behavior for both short and long dynamics. Furthermore, we demonstrated that even when the exact CD potential cannot be constructed, approximate CD terms can still suppress the accumulation of errors. As the quality of the approximation improves, the suppression of the Trotter error becomes increasingly effective. Overall, this work extends the concept of \textit{self-healing}, showing that it can be understood as a manifestation of the underlying oscillatory (harmonic) structure of the Trotter error. 

Here, we showed that the Trotter error of a time-dependent evolution can be decomposed in the adiabatic and digital contribution for all dynamics. Physically speaking, the oscillatory nature of the error can be understood by noting that this framework effectively removes the adiabatic contribution, leaving only the digital contribution. These terms correspond to discrete rotations acting on single and two-qubit states. Intuitively, the repeated application of such discrete rotations, whose angles depend on $\Delta t$ and $\lambda$, naturally leads to an oscillatory behavior of the error. Thus, the terms that damp these oscillations are those associated with the adiabatic contribution to the error. Regarding the regime in which the self-healing mechanism is expected to occur, Fig.~\ref{fig:fit-non-intercting-spin-6}-(d) shows that the ans\"atz proposed in Eq.~\eqref{eq:perturbation-theory-T} fails to accurately describe the dynamics for $T<1$. This can be understood by noting that, in this regime, specific contributions to the error become large. In particular, in the App.~\ref{app:SRQ} in the manuscript, one finds that $M_{ij}$ scales as $1/T$, and therefore grows as $T$ decreases below unity, becoming a non-negligible component of the total error. As a result, the assumptions underlying the ans\"atz break down and the self-healing mechanism is no longer effective. Conversely, for $T>1$ the theoretical description is expected to hold. This discussion is general, i.e. it holds for all evolutions in which we can compute CD potential, so it is also independent on the size $N$ of the system.

From an applied perspective, the results provide actionable guidance for gate-based implementations of digitized state preparation. Once diabatic transitions are sufficiently suppressed (by exact or approximate CD), reducing the Trotter step $\Delta t$ primarily decreases the overall scale of the residual digitization-induced mixing, while increasing $T$ reshapes the phase accumulation that enables destructive interference at the end of the protocol. Endpoint-smooth schedules further favor harmonic cancellation by enforcing boundary conditions on the residual mixing. These considerations suggest a co-design strategy: use the best available diabatic-error suppression to enter the interference-dominated regime, avoid near-resonant windows where digitization costs are amplified, and then allocate circuit depth preferentially to decreasing $\Delta t$.

This work provides a prediction of the error for different values of $\Delta t^2$ and total evolution time $T$ through Eq.~\eqref{eq:perturbation-theory-T}, assuming a linear schedule. From an experimental perspective, one can fix a maximum acceptable error and subsequently determine the corresponding values of $\Delta t$ or $N$, using the relation $T = \lfloor N \Delta t \rfloor$. In the regime where Eq.~\eqref{eq:perturbation-theory-T} can be approximated as scaling like $T^{-2} \Delta t^2$, fixing a target error value $\bar{\mathcal{I}}$ implies that the required number of Trotter steps scales as $\bar{M} \sim 1/\sqrt{\bar{\mathcal{I}}}$. By contrast, according to earlier literature where the error scales as $\Delta t^2 T^2$, one would obtain $\tilde{M} \sim \sqrt{\bar{\mathcal{I}}} / \Delta t^2$. Although the ans\"atz proposed in Eq.~\eqref{eq:perturbation-theory-T} is exact only for specific cases, these arguments indicate that it is nevertheless possible to derive an upper bound for the error. More generally, they show that shortcuts to adiabaticity can effectively limit error accumulation. In the Supplementary Materials, we present an example in which the chosen schedule allows for an analytical solution.

In short, finite-time self-healing emerges as an interference phenomenon intrinsic to digitized dynamics, and counterdiabatic control provides the key lens that exposes---and allows one to exploit---this mechanism beyond the asymptotic adiabatic regime.

\section*{Acknowledgments}

This work is supported by OpenSuperQ+100 (Grant No. 101113946) of the EU Flagship on Quantum Technologies, project Grant No. PID2021-125823NA-I00 funded by MCIN/AEI/10.13039/501100011033 and by ``ERDF A way of making Europe" and ``ERDF Invest in your Future'', and from the IKUR Strategy under the collaboration agreement between Ikerbasque Foundation and BCAM on behalf of the Department of Education of the Basque Government. This project has also received support from the Spanish Ministry of Economic Affairs and Digital Transformation through the QUANTUM ENIA project call - Quantum Spain, and by the EU through the Recovery, Transformation and Resilience Plan -NextGenerationEU and Basque Government through Grant No. IT1470-22, through the ELKARTEK program, project KUBIT (KK-2024/00105).

G. P. and P. L. acknowledge financial support from PNRR MUR Project No. PE0000023-NQSTI. This work was further supported by the European Union’s Horizon 2020 research and innovation program under Grant Agreement No. 101017733, by MUR Project No. CN 00000013-ICSC (P. L.), and by the QuantERA II Programme STAQS project, which received funding from the European Union’s Horizon 2020 research and innovation program under Grant Agreement No. 101017733 (P. L.). X.C. also appreciates 
the Severo Ochoa Centres of Excellence program through Grant CEX2024-001445-S.

We acknowledge the CINECA award under the ISCRA initiative, for the availability of high-performance computing resources and support.

\section*{data availability}
The data that support the findings of this article are publicly available at \cite{vizzuso_data}.

\appendix
\section{Calculation of $\mathcal{S}_{ij}$, $\mathcal{R}_{ij}$ and $\mathcal{Q}_{ij}$}\label{app:SRQ}

To analyze the behavior of the functions $\mathcal{S}_{ij}$, $\mathcal{R}_{ij}$, and $\mathcal{Q}_{ij}$, we work in the instantaneous eigenbasis of the reference (annealing) Hamiltonian
\begin{equation}
H(\lambda)=(1-\lambda)H_{\rm i}+\lambda H_{\rm f},\qquad \lambda\in[0,1],
\end{equation}
defined through
\begin{equation}
H(\lambda)\,|\phi_j(\lambda)\rangle = E_j(\lambda)\,|\phi_j(\lambda)\rangle .
\label{eq:inst-eigenproblem-app}
\end{equation}
We consider stroboscopic times $t_m=m\Delta t$ and $\lambda_m=\lambda(t_m)$, with $\Delta\lambda=\lambda_{m+1}-\lambda_m=\dot{\lambda}(t_m)\Delta t$.

Our goal is to characterize how the error builds up during the digitized evolution. To this end, we consider the overlap amplitudes
$P_i(t_m,\lambda_m)=\langle \phi_i(\lambda_m)|\psi_{\rm T}(t_m)\rangle$ defined in Eq.~\eqref{eq:Pi-definition}.
Using the one-step digitized propagator at fixed $\lambda_m$ (with or without CD, depending on context), the overlaps obey the recursion
\begin{equation}
P_i(t_{m+1},\lambda_{m+1})
=\sum_{j,k} G_{kj}(\lambda_m,\Delta t)\,P_j(t_m,\lambda_m)\,M_{ik}(\lambda_m),
\label{eq:Pi-relation-app}
\end{equation}
where
\begin{equation}
G_{kj}(\lambda,\Delta t)\equiv
\langle\phi_k(\lambda)|U^{\rm CD}_{\rm T}(\lambda,\Delta t)|\phi_j(\lambda)\rangle
\label{eq:Gdef-app}
\end{equation}
captures the action of the (single-step) digitized evolution operator in the instantaneous eigenbasis at fixed $\lambda$, and
\begin{equation}
M_{ik}(\lambda)\equiv \langle \phi_i(\lambda+\Delta\lambda)|\phi_k(\lambda)\rangle
\label{eq:Mdef-app}
\end{equation}
accounts for the basis change between consecutive stroboscopic points. Here $|\psi_{\rm T}(t_m)\rangle$ denotes the \emph{digitized} state, propagated as
$|\psi_{\rm T}(t_{m+1})\rangle = U^{\rm CD}_{\rm T}(\lambda_m,\Delta t)\,|\psi_{\rm T}(t_m)\rangle$.

For sufficiently small $\Delta t$, the diagonal elements satisfy
\begin{equation}
G_{jj}(\lambda,\Delta t)=1-iE_j(\lambda)\Delta t+\mathcal{O}(\Delta t^2).
\end{equation}
Off-diagonal elements arise from product-formula discretization. For first-order Trotterization one has generically
\begin{equation}
G_{i\neq j}(\lambda,\Delta t)
=\mathcal{O}(\Delta t^2)
+\mathcal{O}\!\left(\Delta t^2\,\dot{\lambda}\right),
\label{eq:Goff-app}
\end{equation}
where the second contribution is associated with commutators involving the CD term, e.g.\
$[H_{\rm i,f},\dot{\lambda}\mathcal{A}_\lambda]\sim \mathcal{O}(\dot{\lambda})=\mathcal{O}(1/T)$ for a linear ramp.

On the other hand, expanding the overlap in Eq.~\eqref{eq:Mdef-app} in $\Delta\lambda$ gives
\begin{equation}
M_{ik}=\delta_{ik}
+\Delta \lambda\,\langle \partial_{\lambda}\phi_i|\phi_k\rangle
+\tfrac{1}{2}\Delta \lambda^2\,\langle \partial^2_{\lambda}\phi_i|\phi_k\rangle
+\mathcal{O}(\Delta \lambda^3).
\label{mij_eqn}
\end{equation}
We choose the parallel-transport gauge $\langle \phi_i|\partial_\lambda \phi_i\rangle=0$, so that the linear correction vanishes for $i=k$.
For $i\neq k$, the term $\langle \partial_{\lambda}\phi_i|\phi_k\rangle$ is the usual adiabatic (nonadiabatic) coupling and is generically $\mathcal{O}(1)$, implying
\begin{equation}
M_{i\neq k}(\lambda)=\mathcal{O}(\Delta\lambda)=\mathcal{O}(\Delta t/T)
\qquad\text{(in general)}.
\label{eq:Moff-scaling-app}
\end{equation}

For the {continuous-time} evolution generated by $H_{\rm tot}(t)=H(t)+\dot{\lambda}(t)\mathcal{A}_\lambda$, the CD term cancels the instantaneous nonadiabatic couplings in the adiabatic frame, so that off-diagonal mixing is suppressed at leading order in $\dot{\lambda}$. In the digitized setting, this cancellation is not exact because $U^{\rm CD}_{\rm T} (t)$ is implemented by a product formula; nevertheless, the leading $\mathcal{O}(\Delta t/T)$ mixing originating from the basis change in $M$ is canceled by the CD contribution in $G$ in the $\Delta t\to 0$ limit, leaving residual off-diagonal terms that are higher order in $\Delta t$ and/or $1/T$. This is precisely the origin of the hierarchy discussed below.

Substituting the above expansions into Eq.~\eqref{eq:Pi-relation-app} and collecting terms, one finds three types of contributions:
(i) terms driven by the basis change (adiabatic couplings), which scale as $\sim \Delta t/T$ and are grouped into $\mathcal{S}_{ij}$ and $\mathcal{Q}_{ij}$;
(ii) purely digitization-induced off-diagonal mixing, which survives under CD and is grouped into $\mathcal{R}_{ij}$; and
(iii) higher-order remainders.
In the exact-CD case, the leading $\mathcal{O}(\Delta t/T)$ terms are canceled, so that $\mathcal{S}_{ij}$ and $\mathcal{Q}_{ij}$ are suppressed and the dominant off-diagonal contribution is controlled by $\mathcal{R}_{ij}$.
Keeping the leading terms, the recursion can be written as
\begin{equation}
\begin{split}
P_i(t_{m+1},\lambda_{m+1})
&=\Bigl[1-iE_i(\lambda_m)\Delta t+\mathcal{O}(\Delta t^2)\Bigr]P_i(t_m,\lambda_m) \\
&\quad +\Delta t\sum_{j\neq i}\bar{\mathcal{R}}_{ij}(\lambda_m,\Delta t)\,P_j(t_m,\lambda_m).
\end{split}
\label{eq:P-delta-t}
\end{equation}
Here $\mathcal{O}(\Delta t^2)$ includes subleading $\mathcal{O}(\Delta t^2/T^2)$ terms, and $\bar{\mathcal{R}}_{ij}(\lambda,\Delta t)=\mathcal{O}(\Delta t)$ is the coefficient arising from the off-diagonal part of $G$ (and thus from digitization).
Performing the discrete differentiation $(P_i(t_{m+1})-P_i(t_m))/\Delta t$ yields, to leading order,
\begin{equation}
\begin{split}
i\,\partial_t P_i(t,\lambda)
&\simeq E_i(\lambda)\,P_i(t,\lambda)
+i\sum_{j\neq k}\mathcal{R}_{ij}(\lambda,\Delta t)\,P_j(t,\lambda) \\
&\quad +\mathcal{O}(\Delta t)+\mathcal{O}(\Delta t/T^2).
\end{split}
\label{eq:schrodinger-R}
\end{equation}
where $\mathcal{R}_{ij}$ can be taken of the same order as $\bar{\mathcal{R}}_{ij}$ for the scaling analysis.

Unlike Eq.~\eqref{eq:schrodinger-Pi-complete}, Eq.~\eqref{eq:schrodinger-R} contains only the digitization-induced mixing term. When the evolution is performed {without} the exact CD potential, the cancellation of the $\mathcal{O}(\Delta t/T)$ basis-change couplings does not occur, and the additional terms $\mathcal{S}_{ij}$ and $\mathcal{Q}_{ij}$ must be retained. In that case they scale as $\mathcal{S}_{ij}\sim \mathcal{O}(1/T)$ and $\mathcal{Q}_{ij}\sim \mathcal{O}(1/T)$, recovering the full structure summarized in Eq.~\eqref{eq:schrodinger-Pi-complete}.

The absence of these two terms in Eq.~\eqref{eq:schrodinger-R} is a direct consequence of the fact that the exact dynamics under consideration is adiabatic at all times. Indeed, if the dynamics were not adiabatic, one would obtain
\begin{gather}
    M_{ik}(\lambda)=\braket{\dot{\phi}_i(\lambda)|\phi_j(\lambda)}\Delta t/T+\mathcal{O}((\Delta t/T)^2)\\
    M_{ii}(\lambda)=1+\mathcal{O}(\Delta t/T)
\end{gather}
n this case, all the cross terms arising from Eq.~\eqref{eq:Pi-relation-app} lead to the following definitions of $\mathcal{S}_{ij}$, $\mathcal{Q}_{ij}$, and $\mathcal{R}_{ij}$:
\begin{gather}
G_{jj}(\lambda,\Delta t)M_{jk}(\lambda)\equiv\bar{\mathcal{S}}_{jk}(\lambda,\Delta t)\Delta t+\mathcal{O}(\Delta t^2/T^2)\label{eq:S}\\
G_{jk}(\lambda,\Delta t)M_{jj}(\lambda)\equiv\bar{\mathcal{R}}_{jk}(\lambda,\Delta t)\Delta t+\mathcal{O}(\Delta t^4/T^2)\label{eq:R}\\
\sum_i G_{ji}(\lambda,\Delta t)M_{ik}(\lambda)\equiv\bar{\mathcal{Q}}_{jk}(\lambda,\Delta t)\Delta t+\mathcal{O}(\Delta t^2/T^2).\label{eq:Q}
\end{gather}
In Eq.s~\eqref{eq:S}-~\eqref{eq:R}~\eqref{eq:Q}, the goal is to factor out a term proportional to $\Delta t$. This allows one to differentiate the expression of $P_i(t,\lambda)$ and derive a Schr\"odinger-like equation, analogous to Eq.~\eqref{eq:schrodinger-R}. In this more general case, however, additional contributions to the error appear:
\begin{equation}
\begin{split}
i\,\partial_t P_i(t,\lambda)
&\simeq E_i(\lambda)\,P_i(t,\lambda)
+i\sum_{j\neq k}\Bigl[\mathcal{R}_{ij}+\mathcal{Q}_{ij}+\mathcal{S}_{ij}\Bigr]\,P_j(t,\lambda) \\
&\quad +\mathcal{O}(\Delta t)+\mathcal{O}(\Delta t/T^2).
\end{split}
\label{eq:schrodinger-total}
\end{equation}
As anticipated, the terms $\mathcal{Q}_{ij}$ and $\mathcal{S}_{ij}$ represent the adiabatic contribution to the error and become more significant as $T$ decreases. By introducing a counterdiabatic (CD) potential, these contributions can be eliminated, leaving $\mathcal{R}_{ij}$ as the unique source of error for all values of $T$.

\section{Time dependent perturbation theory}
Eq.~\eqref{eq:schrodinger-Pi-complete} of the main text can be written for $\mathcal{Q}_{ij} = \mathcal{S}_{ij} = 0$ as, 
\begin{align}
i\frac{\partial P_i(t, \lambda)}{\partial t}
\simeq E_i(\lambda)P_i(t, \lambda)
+ i\sum_{i\neq j}\mathcal{R}_{ij}(\Delta t) P_j(t, \lambda),
\label{eq:schrodinger-Pi-complete1}
\end{align}
This can be treated like a Schr\"odinger like equation and can be assumed as a system governed by a Hamiltonian,
\begin{equation}
K (\lambda(t))=K_0(\lambda(t))+V(\lambda(t)); \quad K_0(\lambda)|\phi_i(\lambda)\rangle=E_i(\lambda)|\phi_i(\lambda)\rangle
\end{equation}
Note that, $K_0(\lambda(t))$ is equivalent to the $H(\lambda(t))$ in the main text and $\{\ket{\phi_i}\}$ represents the corresponding eigenbasis. Considering the contributions from the trotter errors are small, we can solve this using time dependent perturbation theory. Let us choose the complete wavefunction as,
\begin{equation}
|\psi(t)\rangle=\sum_i C_i(t)\left[e^{-i \int_0^{t} E_i(\lambda(\tau)) d \tau}\right]|\phi_i(\lambda(t))\rangle.
\end{equation}
Plugging it in the Schr\"odinger equation, we get
\begin{equation}
\dot{C}_i(t)=-i \sum_j V_{ i j}(\lambda(t)) e^{-i \int_0^{t} \Delta_{i j}(\lambda(\tau))  d \tau} C_j(t),
\end{equation}
where $V_{i j}=\langle \phi_i| V|\phi_j\rangle$, $\Delta_{ij}(\lambda(t) )= E_i(\lambda(t)) -E_j(\lambda(t) $. 
From the Dyson expansion we can write down the solution to first order as
\begin{equation}
C_i^{(1)}(t)=-i \int_0^t d t' V_{i j}(\lambda(t') ) e^{-i \int_0^{t'} \Delta_{i j}(\lambda(\tau)) d \tau}
\end{equation}
We can change the integration variable to $t^{\prime} \rightarrow \lambda^{\prime}$, giving $d t^{\prime}=\frac{d \lambda^{\prime}}{\dot{\lambda}(\lambda')}$. A similar change of variable can be done for the integral at the exponent, $\tau \rightarrow \xi$ and $ d\tau =\frac{d \xi}{\dot{\lambda}(\xi)}$, giving us
\begin{equation} 
\label{phink}
\begin{split}
C_i^{(1)}(t)&=-i \int_0^\lambda d \lambda^{\prime} \frac{V_{i j}\left(\lambda^{\prime}\right)}{\dot{\lambda}\left(\lambda^{\prime}\right)} e^{-i \phi_{i j}\left(\lambda^{\prime}\right)} \qquad  \\ \qquad \phi_{i j}\left(\lambda^{\prime}\right)&=\int_0^{\lambda^{\prime}} \frac{\Delta_{i j}(\xi)}{\dot\lambda(\xi)} d \xi.    
\end{split}
\end{equation}
Note that, $C_i^{(1)}(t) \approx P_i(t,\lambda)$ approximated to the first order provided we consider the energy gap between the groundstate and the $i$-th state only 
i.e., $\Delta_{i j}= \Delta_{i0}$. $V_{i j}$ can be chosen as a harmonic perturbation (owing to the boundary conditions of $R_{ij}$) and considering only the leading order $\bar q$: 
$V_{i j}=V e^{i \pi \bar q \lambda}+V^{\dagger} e^{-i \pi \bar q \lambda}$ 
with $V=-i \frac{R}{2}$, which gives
\begin{align}
V_{i j}(t) =R \sin (\pi \bar q \lambda).
\end{align}
Therefore equation \eqref{phink} can be re written as
\begin{equation}
\label{eq:constdel}
\begin{split}
P_i( t, \lambda)&=-i \int_0^\lambda d \lambda^{\prime} \frac{R \sin \left(\pi \bar q \lambda^{\prime}\right)}{\dot{\lambda}\left(\lambda^{\prime}\right)} e^{-i \varphi} \qquad \\ \qquad \varphi&=\int_0^{\lambda'} \frac{\Delta_{i0}}{\dot{\lambda}(\xi)} d \xi.
\end{split}
\end{equation}

\section{Solution for $\lambda(t) = \sin^2\left(\frac{\pi t}{2 T}\right)$}

Eq.~\eqref{eq:constdel} is highly non  trivial in terms of $\lambda$ and difficult to solve for complex schedule function. furthermore, as described in the main text, the choice of the schedule function in a way such that the boundary condition for the CD potential has to be satisfied, that is, $\left.\lambda\right|_{t=0}=0,\left.\lambda\right|_{t= T}=1$ and also, $ \left.\dot\lambda\right|_{t=0} = \left.\dot\lambda\right|_{t=T} =0 $. One such choice is the following,
\begin{equation}
\lambda=\sin ^2 \left(\frac{\pi t}{2 T}\right) \quad \Rightarrow \quad \dot\lambda=\frac{\pi}{T} \sqrt{\lambda(1-\lambda)}.
\end{equation}
With this choice we can express Eq.~\eqref{eq:constdel} as
\begin{equation}
P_i( T,1)  =-\frac{i T R}{\pi} \int_{0}^{1} d \lambda^{\prime} \frac{\sin \left(\bar q \pi \lambda^{\prime}\right)}{\sqrt{\lambda^{\prime}\left(1-\lambda^{\prime}\right)}} e^{-\frac{i 2 T \Delta_{i0}}{\pi} \sin ^{-1} \sqrt{\lambda^{\prime}}}
\end{equation}
This equation can further be simplified with the substitution $\lambda^{\prime}=\sin ^2 \theta$ for $\theta=\{0, \pi / 2\} $,
\begin{equation}\label{thetaeqnfinal}
\begin{split}
    P_i( T,1)&=-\frac{2 i T R}{\pi} \int_0^{\pi / 2} d \theta \sin \left(\bar q \pi \sin ^2 \theta\right) e^{-i \kappa \theta} d \theta \quad \\ &\text{with}\quad \kappa=\frac{2 T \Delta_{i0}}{\pi}.    
\end{split}
\end{equation}
We can find an analytical solution for Eq.~\eqref{thetaeqnfinal} using the Jacobi-Anger expansion, given by 
\begin{equation} \label{jacobi}
e^{i z \cos \zeta}=\sum_{n=-\infty}^\infty i^n J_n(z) e^{i n \zeta},
\end{equation}
where $J_n(z)$ are the Bessel function of first kind. Eq.~\eqref{jacobi} can be divided into real and imaginary parts, and for our case,
\begin{align}
\cos (\alpha \cos 2\theta)  &= J_0(\alpha)+2 \sum_{n=1}^\infty(-1)^n J_{2 n}(\alpha) \cos(4 n \theta), \qquad \text{even}\\
\sin (\alpha \cos 2\theta) &= 2 \sum_{n=1}^{\infty}(-1)^n J_{2 n+1}(\alpha) \cos((4 n+2) \theta), \qquad \text{odd}
\end{align}
where we are using the identity $\sin \left(m \pi \sin ^2 \theta\right) = \sin \alpha \cos (\alpha \cos 2 \theta)-\cos \alpha \sin (\alpha \cos 2 \theta)$ with $ \alpha=\frac{\bar q \pi}{2}$.
From, Eq.~\eqref{thetaeqnfinal} it is straightforward to write,
\begin{align}
& \int_0^{\pi / 2} \cos (\alpha \cos 2 \theta)e^{-i k \theta} d \theta \\
= & \int_0^{\pi / 2}\left[J_0(\alpha)+2 \sum_{n=1}^\infty(-1)^n J_{2 n}(\alpha) \cos (4 n \theta)\right] e^{-i \kappa \theta} d \theta \\
= & \int_0^{\pi / 2} J_0(\alpha) e^{-i \kappa \theta} d \theta+\\ &2 \sum_{n=1}^\infty(-1)^n J_{2 n}(\alpha) \int_0^{\pi / 2} \cos (4 n \theta) e^{-i \kappa \theta} d \theta.
\end{align}
Defining the standard integrals like
\begin{align}
    I_0(\kappa) &:= \int_0^{\pi / 2} e^{-i \kappa \theta} d \theta=\frac{1-e^{-i \kappa \pi / 2}}{i \kappa}\\
    I_p(\kappa) &:= \int_0^{\pi / 2} \cos (p \theta) \,  e^{-i \kappa \theta} d \theta =\\& = \frac{1}{2}\left[\frac{1-e^{-i(\kappa-p) \pi / 2}}{i(\kappa-p)}+\frac{1-e^{-i(\kappa+p) \pi / 2}}{i(\kappa+p)}\right],
\end{align}
we can write the solution as
\begin{equation}
\begin{split}
\int_0^{\pi / 2} \cos (\alpha \cos 2 \theta)e^{-i k \theta} d \theta &\\= I_0(k) J_0(\alpha)+2 \sum_{n=1}^\infty (-1)^n J_{2 n}(\alpha) &I_{4 n}(k)
\end{split}
\end{equation}
and, similarly,
\begin{equation}
\int_0^{\pi / 2} \sin (\alpha \cos 2 \theta)e^{-i k \theta} d \theta = 2 \sum_{n=1}^{\infty} (-1)^n J_{2 n+1}(\alpha) I_{4 n+2}(k).
\end{equation}
Finally we can write the full analytical solution as
\begin{equation} \label{final}
\begin{split}
P_i(T, 1)&=-\frac{i 2 T R}{\pi}\left[I_0(k) J_0(\alpha)+2 \sum_{n=1}^\infty (-1)^n J_{2 n}(\alpha) I_{4 n}(k)\right]\\&-\frac{i 2 T R}{\pi}\left[2 \sum_{n=1}^{\infty} (-1)^n J_{2 n+1}(\alpha) I_{4 n+2}(k)\right]
\end{split}
\end{equation}
with $\alpha=\frac{\bar q \pi}{2}$, $k=\frac{2 T\Delta_{i0}}{\pi}$.

\begin{figure}
\centering
    \includegraphics[width=0.5\textwidth]{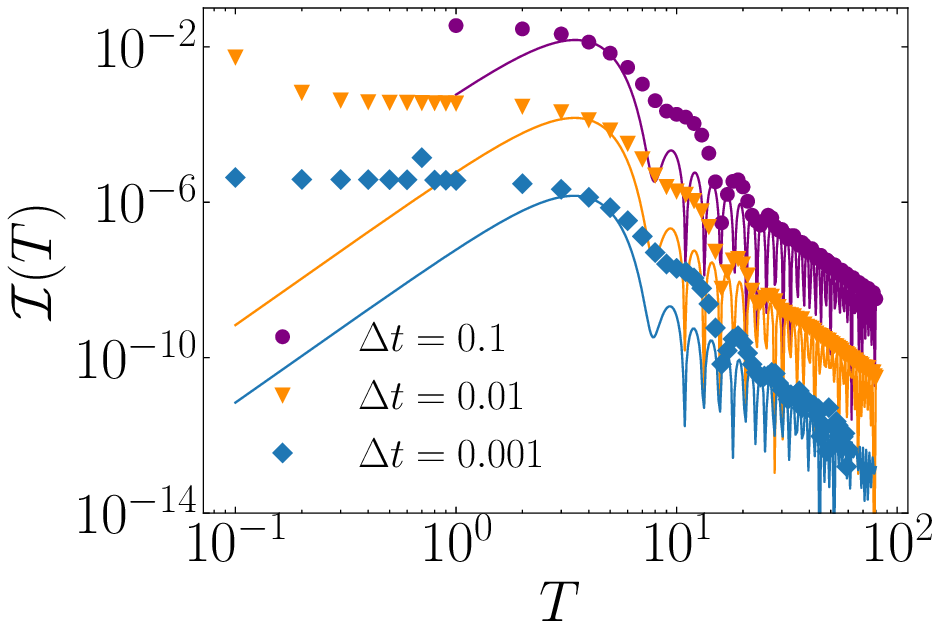}
    \caption{Infidelity versus the total evolution time $T$ for the single qubit model, shown for three Trotter step sizes: $\Delta t = 0.1$ (purple), $\Delta t = 0.01$ (yellow), and $\Delta t = 0.001$ (blue). The infidelity data are fitted using the leading–order term of Eq.~\eqref{final} with $\bar q = 2$, yielding the fitted values $\Delta = 2.579,\, 2.596,\, 2.611$ for the three cases, respectively. 
    }
    \label{fig:bessel_fit_non-interacting}
\end{figure}

For validation, we fit the function in Eq.~\eqref{final} for the single qubit model. As shown in Fig.~\ref{fig:bessel_fit_non-interacting}, the fit aligns very well with the numerical data when using the leading mode $\bar q = 2$, analogous to the linear-schedule case. The corresponding fitted energy gaps are found to be $\Delta_{i0} = 2.579,\, 2.596,\, 2.611$ for $\Delta t = 0.1,\, 0.01,\,\text{and } 0.001$, respectively.
Note that this fit is not fully optimal, since only a single mode is included; a more accurate representation would require incorporating several modes. Nevertheless, the single-mode fit already provides a clear indication of the convergence behavior of the Trotter error. In the large-$T$ limit, the scaling follows the expected $\mathcal{O}(1/T^{2})$, while for intermediate $T$ the convergence is well captured by the fitting function. However, for $T < 1$, 
the approximation breaks down as higher-order terms begin to contribute 
significantly.

\section{Boundary condition for the linear ramp}

For the linear ramp schedule, the function $\mathcal{R}_{ij}(t)$ cannot be treated as a harmonic function, since we cannot assume that $\mathcal{R}_{ij}(t=0)=\mathcal{R}_{ij}(t=T)=0$ and $ \dot\lambda({t=0}) = \dot\lambda({t=T}) \neq 0 $. This introduces an offset in the usual harmonic perturbation. To rectify this let us denote the offset by $\mathcal{R}_{ij}(0)\equiv a$ and $\mathcal{R}_{ij}(T)\equiv b$ and define a new function
\begin{equation}
    r_{ij}(t)=\mathcal{R}_{ij}(t)-a+(a-b)\frac{t}{T},
    \label{eq:r}
\end{equation}
which satisfies $r_{ij}(0)=r_{ij}(T)=0$ and can therefore be expanded as a harmonic function. Thus, in this case the error $\mathcal{R}_{ij}(t)$ can be decomposed into a linear part and a harmonic part. Consequently, the Trotter error can be interpreted as arising from the action of an effective potential of the form
\begin{equation}
    V_{ij}(\Delta t,\lambda)
    = a + (b-a)\frac{t}{T}
    + V e^{i\pi \bar{q}t/T}
    + V^\dagger e^{-i\pi \bar{q}t/T}.
    \label{eq:linear-harmonic-perturabetion}
\end{equation}
Using Eq.~\eqref{eq:linear-harmonic-perturabetion} for $V(t)$ at $\lambda=1$,  
we compute
\begin{equation}
\begin{split}
    P_i(t, \lambda) = &\\
    \int_{0}^t dt'\!
    \left(a+(b-a)\frac{t'}{T}\right)e^{-i\Delta_{nm}t'}
    \nonumber
    \\- \int_{0}^t dt'\frac{iR}{2}
    \left(e^{i\pi\bar{q}t'/T}+e^{-i\pi\bar{q}t'/T}\right)
    e^{-i\Delta_{nm}t'},
\end{split}
    \label{eq:integral}
\end{equation}
This is a general formula that we used for Eq.~\eqref{eq:constdel}.  The solution for $t=T$ can therefore be written as, 

\begin{equation}
\begin{split}
    P_i(T,1) =& \frac{ia}{\Delta_{i0}}
    \left(e^{-iT\Delta_{i0}}-1\right)
    \\+\frac{(b-a)}{\Delta_{i0}}&
    \left(
        iT e^{-iT\Delta_{i0}}
        +\frac{1}{\Delta_{i0}T}
        \left(e^{-iT\Delta_{i0}}-1\right)
    \right)
    \\+\frac{R}{2}&
    \left((-1)^{\bar{q}}e^{-i\Delta_{i0}T}-1\right)
    \frac{\bar{q}\pi}{
        \Delta_{i0}T-\frac{\pi^2\bar{q}^2}{T}
    }.
\end{split}
    \label{eq:perturbatio-theory-T-final}
\end{equation}
However, estimating $a$ and $b$ for general many-body systems is difficult, and fitting both parameters is equally challenging. 
In contrast, for non-interacting systems these quantities can be computed analytically. 
In such systems the only relevant CD matrix elements are $\mathcal{R}_{01}(t)$ and $\mathcal{R}_{10}(t)$, since the dynamics reduces to an effective two-level problem. Moreover, because the operators involved are Hermitian and we ultimately consider absolute values, it is sufficient to focus on a single term, namely $\mathcal{R}_{01}(t)$. 
In this case we have
\begin{equation}
    \begin{split}
        \mathcal{R}_{01}(0)
        &= \braket{+|e^{-iH_\text{CD}(0)\,\Delta t}|-}, \\
        \mathcal{R}_{01}(T)
        &= \braket{\uparrow|e^{-iH_\text{CD}(T)\,\Delta t}|\downarrow},
    \end{split}
    \label{eq:a-b-one-body}
\end{equation}
where $\ket{+},\ket{-}$ and $\ket{\uparrow},\ket{\downarrow}$ are the eigenstates of the initial and final Hamiltonians, respectively.

For a single qubit we have~\cite{Passarelli2020}
\begin{equation}
    H_\text{CD}(t)=i\alpha\sigma_y,
    \qquad
    \alpha=-\frac{1}{4-8\lambda(t)+8\lambda^2(t)},
\end{equation}
where $\sigma_y$ is the Pauli-$y$ matrix. 
Thus, since $\dot\lambda$ is constant, during all the evolution and in particular at $\lambda(t)=0,1$ we have
\begin{equation}
    H_\text{CD}(0)=H_\text{CD}(T)
    =-\frac{1}{4T}\sigma_y,
    \label{eq:CD-boundary-one-body}
\end{equation}
leading to
\begin{equation}
    a=b=-\Delta t\,
    \sin\!\left(\frac{1}{4T}\right).
    \label{eq:R-boundary-one-spin}
\end{equation}

Substituting into Eq.~\eqref{eq:perturbatio-theory-T-final} gives
\begin{equation}
\begin{split}
    P_i(T,1) = &\\\frac{-i\Delta t\,\sin(1/4T)}{\Delta_{i0}}
    &\left(e^{-iT\Delta_{i0}}-1\right)
    \\+\frac{R}{2}
    \left((-1)^{\bar{q}}e^{-i\Delta_{i0}T}-1\right)&
    \frac{\bar{q}\pi}{
        \Delta_{i0}T-\frac{\pi^2\bar{q}^2}{T}
    }.
\end{split}
    \label{eq:Pi-one-body}
\end{equation}

\begin{figure}
    \centering
    \includegraphics[width=1\linewidth]{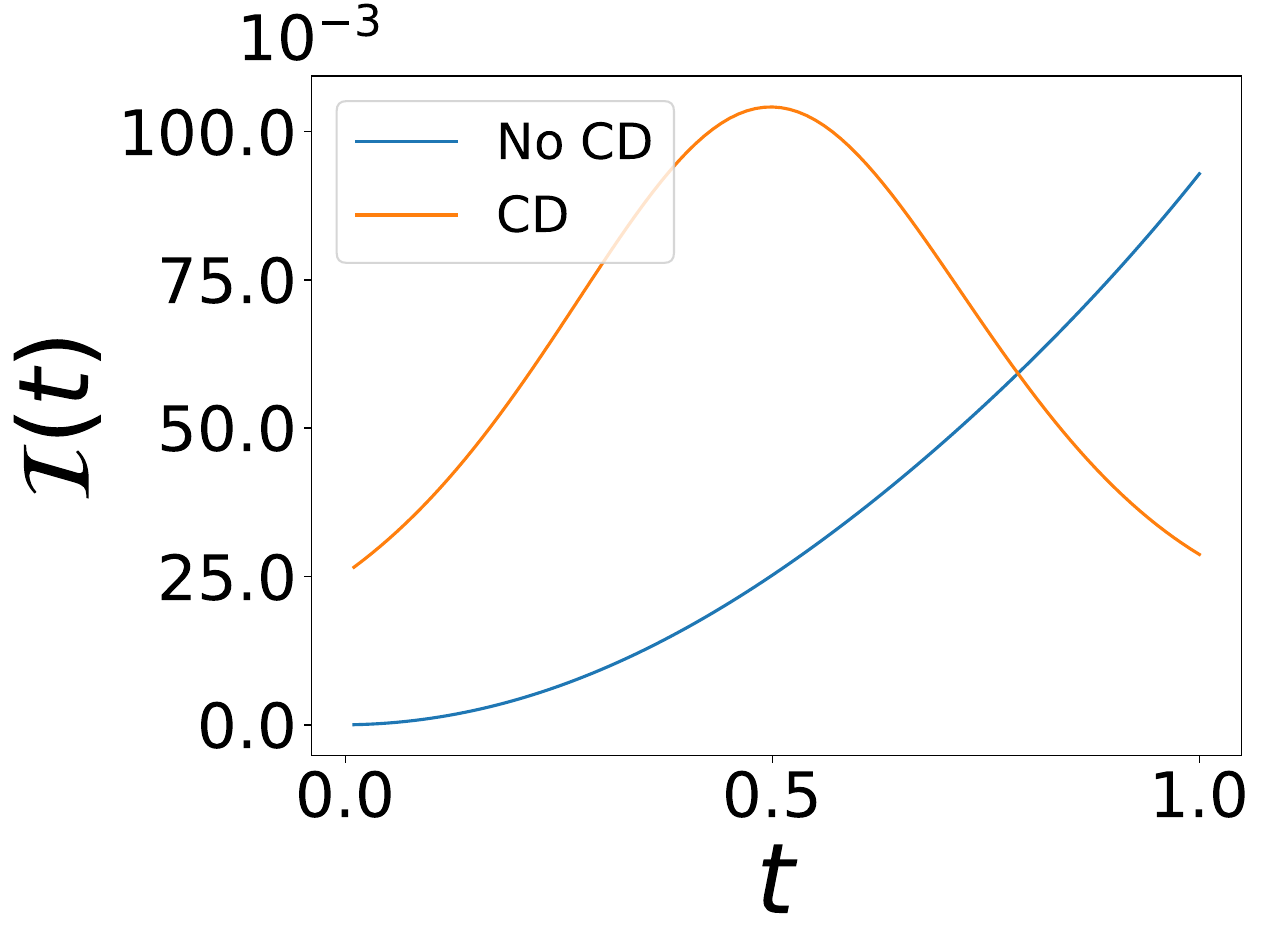}
    \caption{Dynamics of infidelity for the single qubit model, $T=1$, $\Delta t=0.01$, with and without the counterdiabatic correction, in the case of a linear ramp with uncorrected initial condition.}
    \label{fig:one-body-ramp-no-bound}
\end{figure}

Eq.~\eqref{eq:Pi-one-body} shows that the addition of the counterdiabatic potential does not guarantee the absence of a constant contribution in the error: the perturbation can in fact be split into two components, a constant term $a$, independent of $t$, and a harmonic function. Apart from the constant offset $a$, the remaining contribution is harmonic. 

A natural question is whether it is possible to eliminate the constant term $a$ and make the expression in Eq.~\eqref{eq:Pi-one-body} purely harmonic. The answer is yes: this can be enforced by imposing the initial-state condition $P_i(t=0,\lambda=0) \equiv 0$. Indeed, we observe that  
\begin{equation}
    P_1(t=0,\lambda=0)
    = \braket{\phi_1(\lambda=0)\,|\, e^{-iH_{\mathrm{CD}}(0)} |\psi_0 }
    = \mathcal{R}_{01}(0)
    = a,
    \label{eq:initial-condition}
\end{equation}
so that requiring $P_i(t=0,\lambda=0)\equiv 0$ directly implies $a \equiv 0$. Under this condition, the theoretical error is entirely captured by the harmonic term in Eq.~\eqref{eq:Pi-one-body}. 

The same discussion can be extended to interacting models. In that case, however, the coefficients $a$ and $b$ are not a priori equal. Nevertheless, since both contributions scale as $\sim \Delta t$, their difference $a-b$ can be regarded as negligible. 
In Fig.~\ref{fig:one-body-ramp-no-bound} we illustrate what happens when the initial condition is left unmodified for the ramp schedule. Throughout the entire evolution, a finite offset persists, of the order of $H_{\mathrm{CD}}(t=0)$.

In summary, by ensuring that the constant term $a$ is set to zero, the error for the ramp protocol can always be upper bounded by  
\begin{equation}
    \mathcal{I}(T) \le |P_i(T,1)|^2
    = \left|
        \frac{R}{2}
        \left( (-1)^{\bar q} e^{-i \Delta_{i0} T} - 1 \right)
        \frac{\bar q \pi}{
            \Delta_{i0} T - \dfrac{\pi^2 \bar q^2}{T}
        }
    \right|^2,
    \label{eq:Pi-ramp}
\end{equation}
where we retain a single leading-order contribution labeled by $\bar q$, and introduce an effective energy gap $\Delta_{i0}$ associated with an effective level $i$.

\section{Additional numerical results}\label{extra_results}

\begin{figure}
\centering
    \hspace{0.1\columnwidth}(a)\hspace{0.4\columnwidth}(b)\\
    \includegraphics[width=0.46\columnwidth]{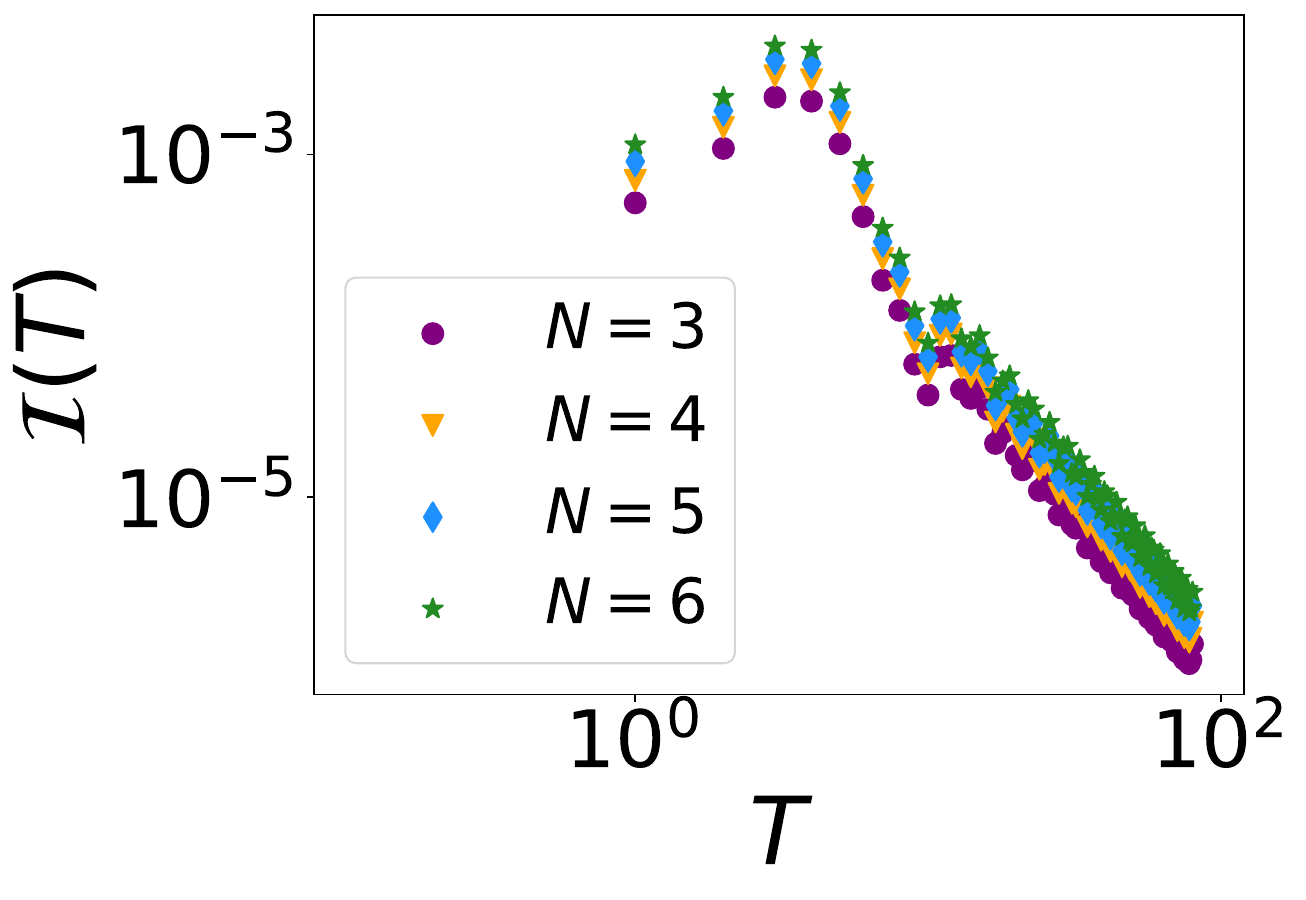}
    \includegraphics[width=0.46\columnwidth]{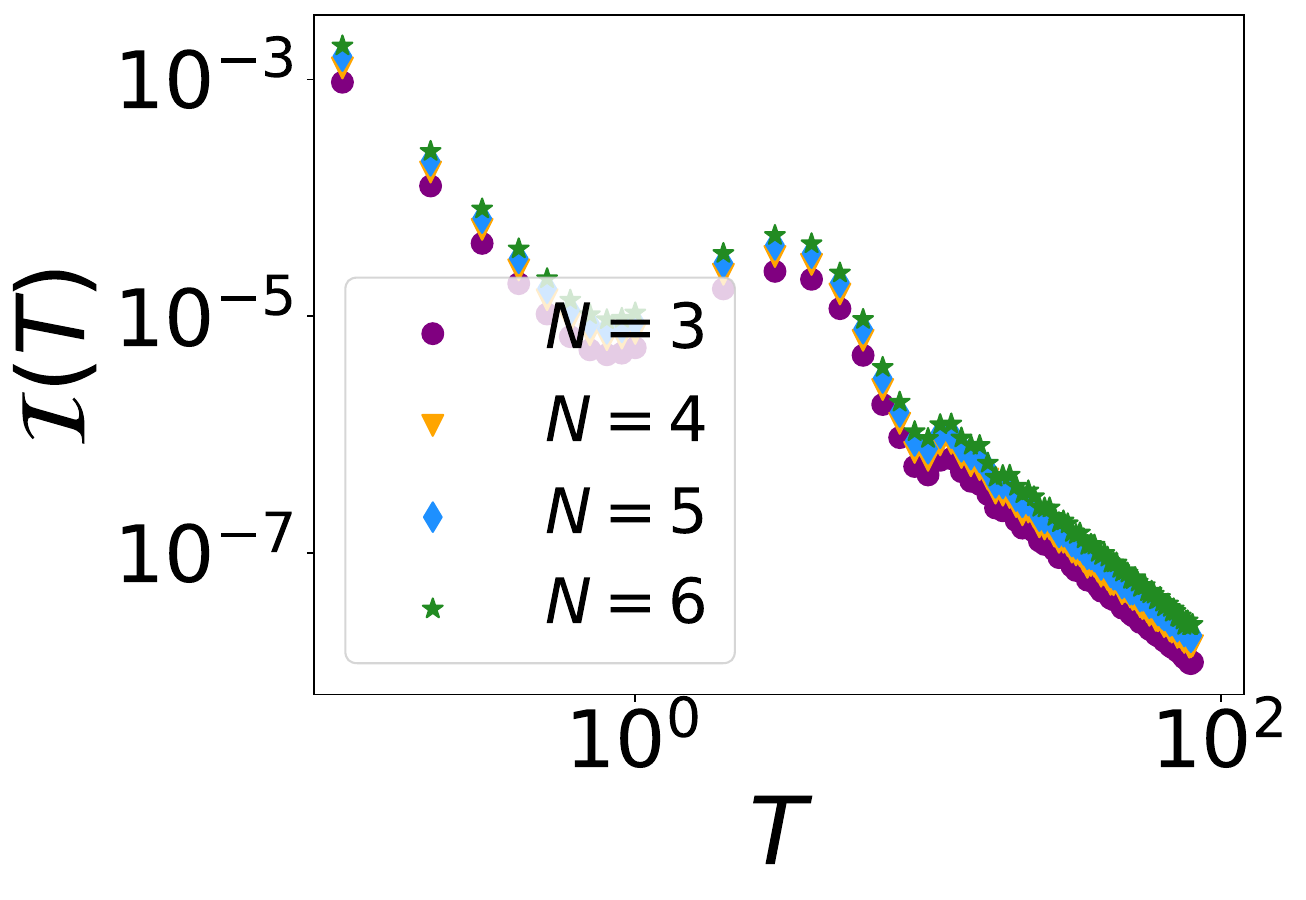}\\
    \hspace{0.1\columnwidth}(c)\hspace{0.4\columnwidth}(d)\\
    \includegraphics[width=0.46\columnwidth]{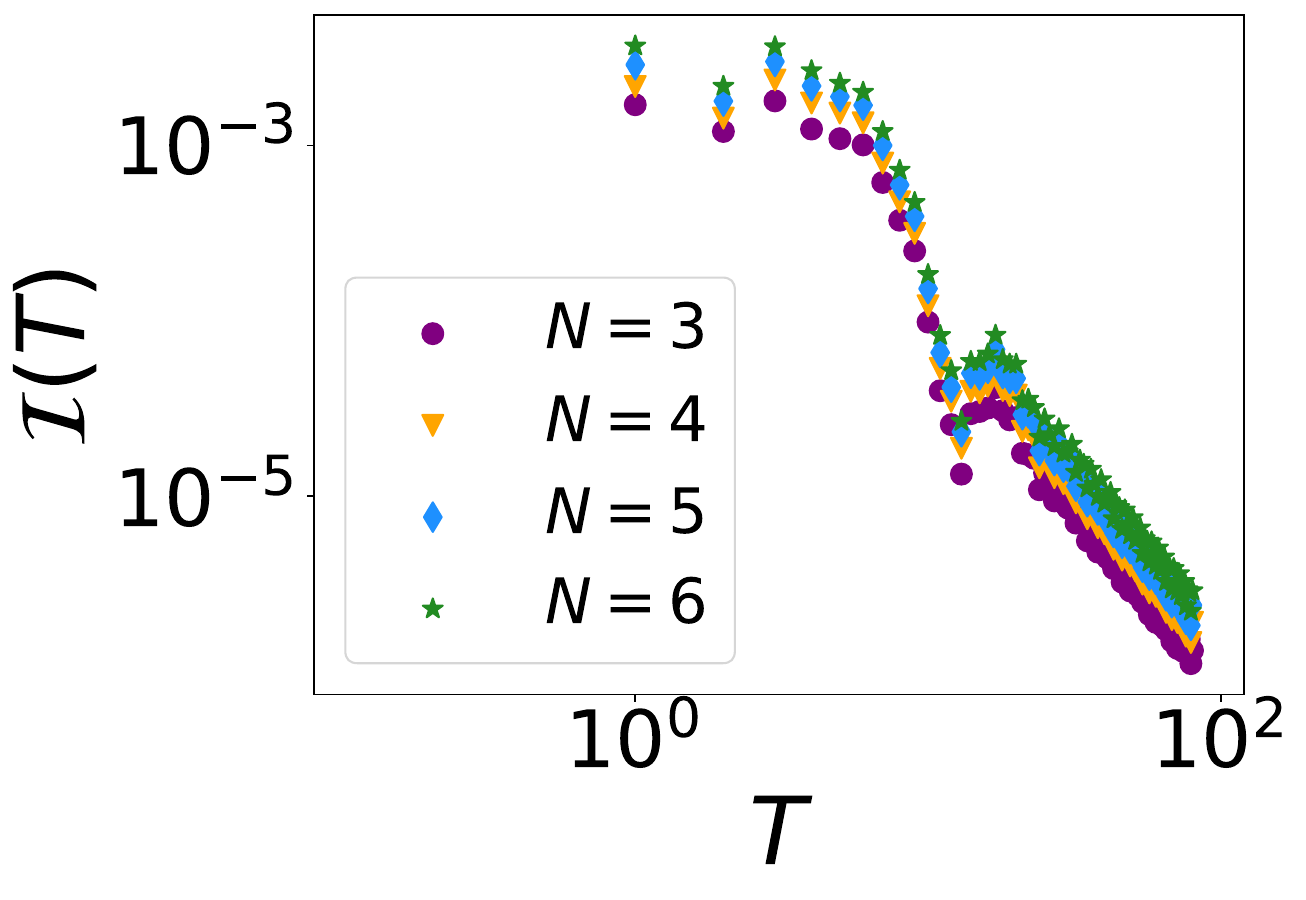}
    \includegraphics[width=0.46\columnwidth]{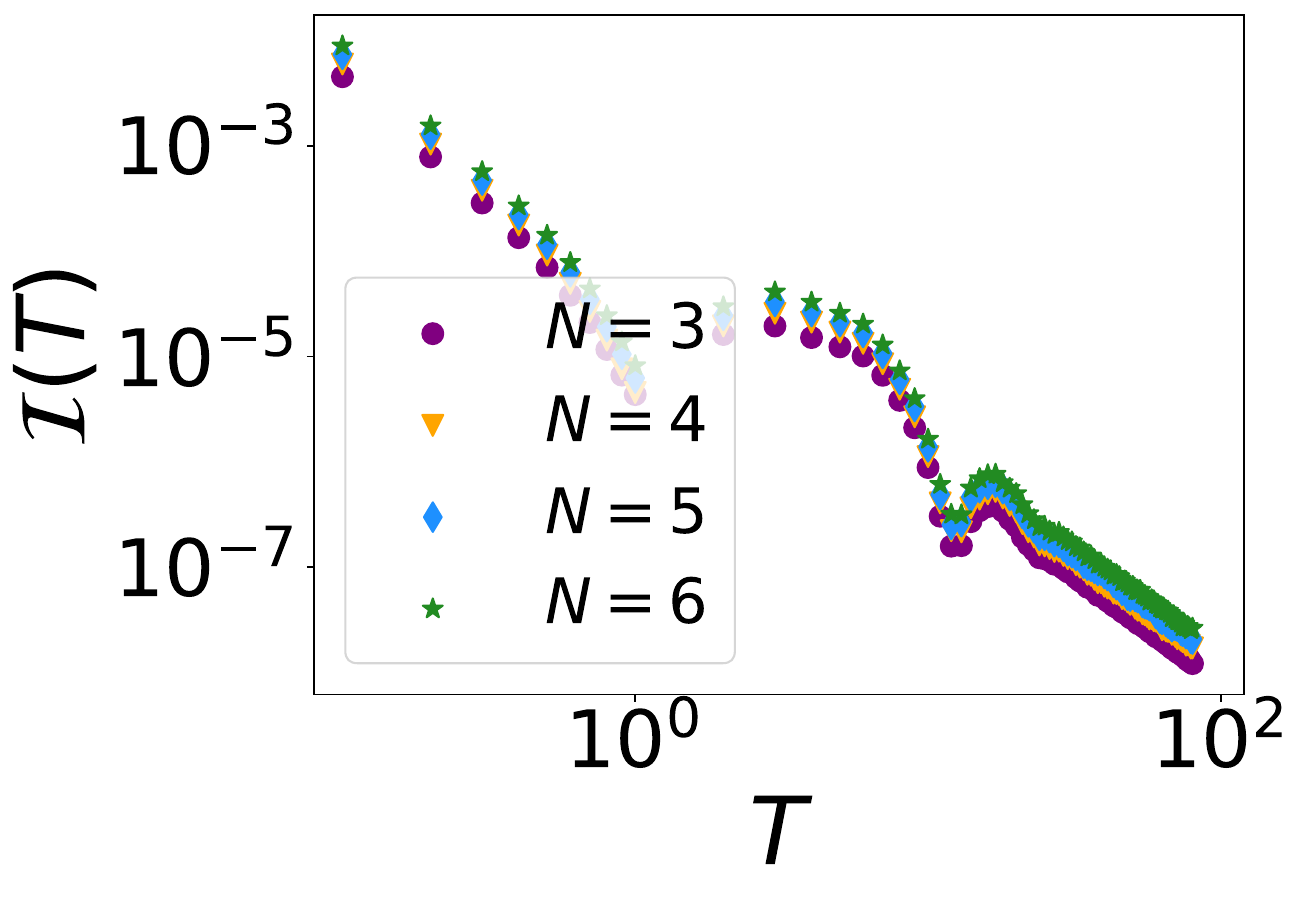}\\
    \hspace{0.1\columnwidth}(e)\hspace{0.4\columnwidth}(f)\\
    \includegraphics[width=0.46\columnwidth]{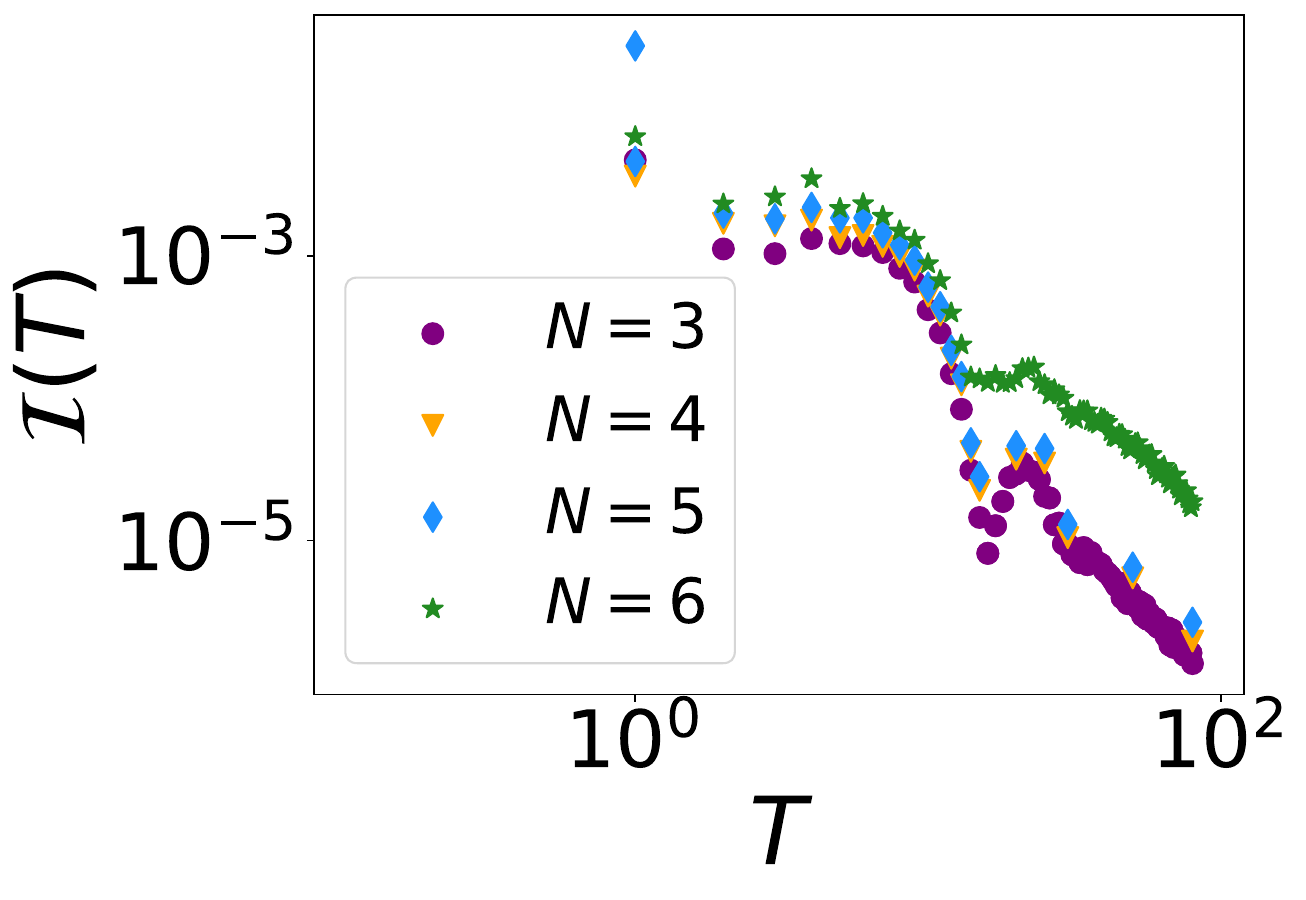}
    \includegraphics[width=0.46\columnwidth]{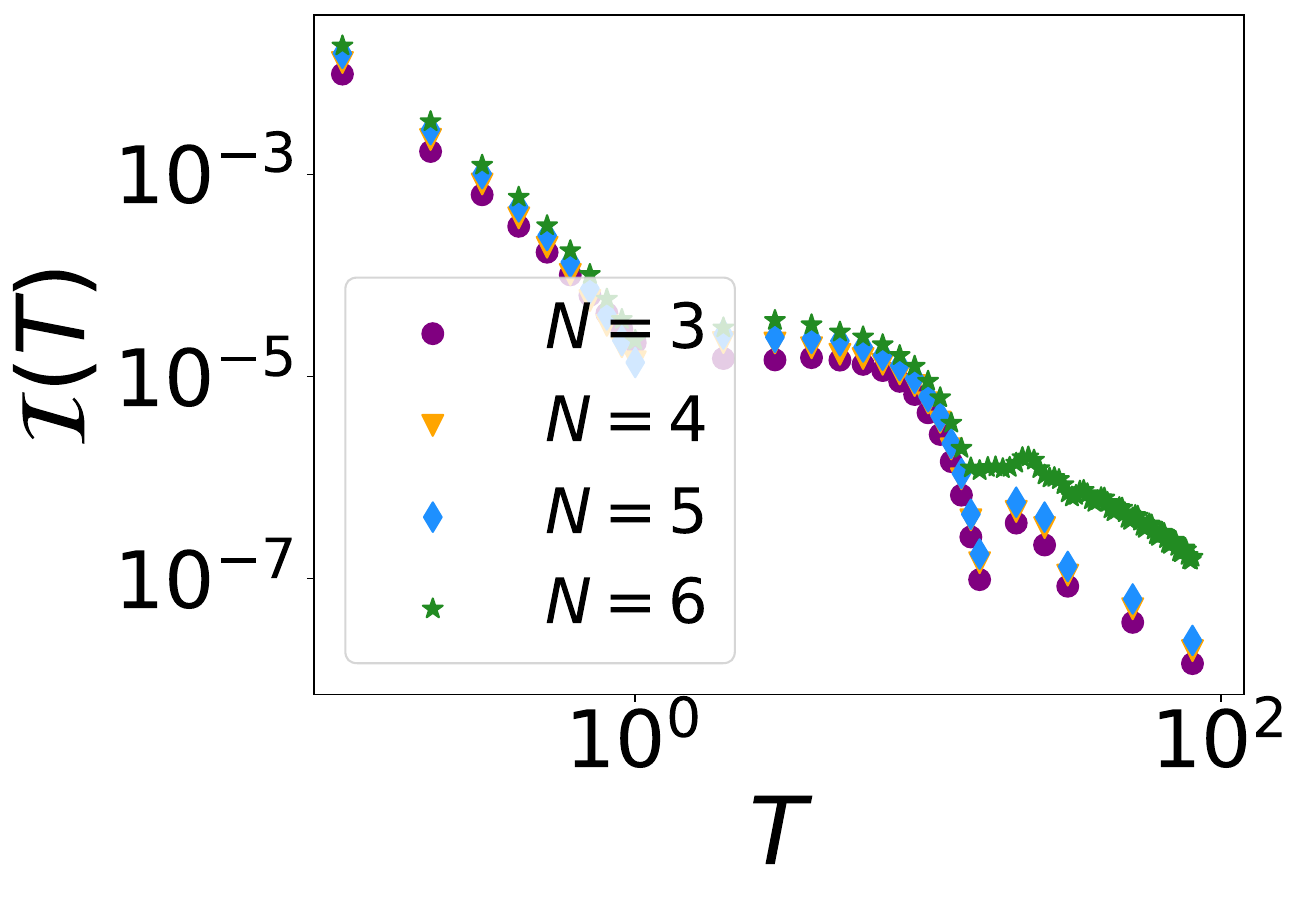}
    \caption{Infidelity vs evolution time $T$ for the Ising chain, for different combination of parameters and system sizes. In panels a)-c)-e) $\Delta t=0.1$ and in panels b)-d)-f) $\Delta t = 0.01$. The final Hamiltonian is Eq.~\eqref{eq:hz} for $J_Z=0.1$ (a)-b)), for $J_Z = 0.5$ (c)-d)) and $J_Z=1$ (e)-f)) and for different sizes. The evolution is given by Eq.~\eqref{eq:CD-real-potential}. Violet points represent system of $3$ spins, blue represents $4$ spins, black is for $5$ spins and yellow for $6$ spins.}
    \label{fig:JZ-0.1-0.5-1-ising-model}
\end{figure}

For completeness, in Fig.~\ref{fig:JZ-0.1-0.5-1-ising-model} we report the infidelity $\mathcal{I}(T)$ at the final time $T$ for the Ising model evolved with the exact CD potential for the ramp schedule. Violet points correspond to a spin size of $N=3$, yellow to $N=4$, blue to $N=5$, and green to $N=6$. The left panels show the results for $\Delta t = 0.1$, while the right panels correspond to $\Delta t = 0.01$. The upper panels refer to $J_Z = 0.1$, the middle ones to $J_Z = 0.5$, and the bottom ones to $J_Z = 1$.

%

\end{document}